\documentclass[aps,prd,twocolumn,preprintnumbers,nofootinbib,10pt]{revtex4-1}%

\pdfoutput=1
\usepackage{amsmath,amsthm,amssymb,multirow,psfrag}
\usepackage{epsfig}
\usepackage{color}

\graphicspath{{./Figures/}}

\begin{document}

\def\lsim{\mathrel{\rlap{\lower4pt\hbox{\hskip1pt$\sim$}}F
  \raise1pt\hbox{$<$}}}
\def\gsim{\mathrel{\rlap{\lower4pt\hbox{\hskip1pt$\sim$}}
  \raise1pt\hbox{$>$}}}
\newcommand{\vev}[1]{ \left\langle {#1} \right\rangle }
\newcommand{\bra}[1]{ \langle {#1} | }
\newcommand{\ket}[1]{ | {#1} \rangle }
\newcommand{\ev}{ {\rm eV} }
\newcommand{\kev}{{\rm keV}}
\newcommand{\mev}{{\rm MeV}}
\newcommand{\tev}{{\rm TeV}}
\newcommand{\mpl}{$M_{Pl}$}
\newcommand{\mw}{$M_{W}$}
\newcommand{\Ft}{F_{T}}
\newcommand{\Zparity}{\mathbb{Z}_2}
\newcommand{\BLambda}{\boldsymbol{\lambda}}
\newcommand{\met}{\;\not\!\!\!{E}_T}

\newcommand{\beq}{\begin{equation}}
\newcommand{\eeq}{\end{equation}}
\newcommand{\bea}{\begin{eqnarray}}
\newcommand{\eea}{\end{eqnarray}}
\newcommand{\nn}{\nonumber \\ }
\newcommand{\gev}{{\mathrm GeV}}
\newcommand{\hc}{\mathrm{h.c.}}
\newcommand{\eps}{\epsilon}

\newcommand{\cO}{{\cal O}}
\newcommand{\cL}{{\cal L}}
\newcommand{\cM}{{\cal M}}

\graphicspath{{./Figures/}}

\newcommand{\fref}[1]{Fig.~\ref{fig:#1}} 
\newcommand{\eref}[1]{Eq.~\eqref{eq:#1}} 
\newcommand{\aref}[1]{Appendix~\ref{app:#1}}
\newcommand{\sref}[1]{Sec.~\ref{sec:#1}}
\newcommand{\tref}[1]{Table~\ref{tab:#1}}  

\def\TY#1{{\bf  \textcolor{red}{[TY: {#1}]}}}
\newcommand{\draftnote}[1]{{\bf\color{blue} #1}}
\newcommand{\draftnoteR}[1]{{\bf\color{red} #1}}

\title{ {\bf \Large{Supersymmetric Custodial Higgs Triplets\\ and the Breaking of Universality}\normalsize}}
\author{\sc{Mateo Garcia-Pepin$\,^{a}$,~Stefania Gori$\,^{b}$,~Mariano Quiros$\,^{a,c}$,\\~Roberto Vega$\,^{d}$,~Roberto Vega-Morales$\,^{e}$,~Tien-Tien Yu$\,^{f}$}\\~\\}

\affiliation{
$^a$IFAE-Institut de Fisica d'Altes Energies, Universitat Aut$\grave{o}$noma de Barcelona, 08193 Bellaterra, Barcelona, Spain\\
$^b$Perimeter Institute for Theoretical Physics, 31 Caroline St. North, Waterloo, Ontario, Canada\\
$^c$ICREA-Instituci$\acute{o}$ Catalana de Recerca i Estudis Avan\c{c}ats, 08015 Barcelona, Spain\\
$^d$Department of Physics, Southern Methodist University, Dallas, TX 75275, USA\\
$^e$Laboratoire de Physique Th\'{e}orique, CNRS - UMR 8627, Universit\'{e} Paris-Sud, Orsay, France\\
$^f$C.N. Yang Institute for Theoretical Physics, Stony Brook University, Stony Brook, NY 11794, USA }

\begin{abstract}
Higgs triplet models are known to have difficulties obtaining agreement with electroweak precision data and in particular constraints on the $\rho$ parameter.~Either a global $SU(2)_L \otimes SU(2)_R$ symmetry has to be imposed on the scalar potential at the electroweak scale, as done in the well-known Georgi-Machacek (GM) model, or the triplet vacuum expectation values must be very small.~We construct a supersymmetric model that can satisfy constraints on the $\rho$ parameter, even if these two conditions are not fulfilled.~We supersymmetrize the GM model by imposing the $SU(2)_L \otimes SU(2)_R$ symmetry at a scale $\mathcal M$, which we argue should be at or above the messenger scale, where supersymmetry breaking is transmitted to the observable sector.~We show that scales $\mathcal M$ well above 100 TeV and sizable contributions from the triplets to electroweak symmetry breaking can be comfortably accommodated.~We discuss the main phenomenological properties of the model and demonstrate that the departure from custodial symmetry at the electroweak scale, due to radiative breaking, can show up at the LHC as a deviation in the `universal' relation for the Higgs couplings to $WW$ and $ZZ$.~As a by-product of supersymmetry, we also show that one can easily obtain both large tree-level \emph{and} one loop corrections to the Higgs mass.~This allows for top squarks that can be significantly lighter and with smaller mixing than those needed in the MSSM.
\end{abstract}

\preprint{UAB-FT-761,~LPT-Orsay-14-71,~YITP-SB-14-34,~SMU-HEP-14-07}

\maketitle

\section{Introduction}\label{sec:Intro}
Establishing the precise nature of the mechanism responsible for electroweak symmetry breaking (EWSB) is one of the primary theoretical goals in particle physics and one of the main objectives of the LHC.~The recent discovery of a resonance with mass $\sim 125$~GeV by the ATLAS and CMS~\cite{Chatrchyan:2012ufa,Aad:2012tfa} collaborations, which has couplings to gauge bosons similar to those of the Standard Model (SM) Higgs~\cite{Falkowski:2013dza}, seems to point towards the Higgs $SU(2)_L$ doublet structure of the SM.~However, well-known theoretical arguments lead to the suspicion that this discovery is not yet the full story for EWSB and, furthermore, the uncertainties in the experimental determination~\cite{CMS-PAS-HIG-14-009,ATLAS-CONF-2014-009} of the Higgs properties still leave room for extended Higgs sectors which might contribute to EWSB and the $W$ and $Z$ boson masses.

One of the simplest extensions of the SM Higgs sector which can contribute to EWSB consists of an additional $SU(2)_L$ doublet, as in the Minimal Supersymmetric Standard Model (MSSM).~This is the smallest single irreducible representation of $SU(2)_L \times U(1)_Y$ satisfying the necessary condition~\cite{Low:2010jp} for preserving the well-known custodial symmetry of the gauge boson mass matrix, typically associated with $\rho = 1$ at tree-level.~An important feature of this representation is that  the condition $\rho_{tree} = 1$ is satisfied even if the vacuum expectation values (VEVs) of the two doublets are misaligned ($\tan\beta\neq 1$) and therefore, even in the case of custodial symmetry breaking.~Note that for these $SU(2)_L \times U(1)_Y$ representations the conditions which gives $\rho_{tree} =1$ imply the well-known tree-level `universality' relation~\cite{Low:2010jp} for the $WW$ and $ZZ$ couplings $g_{\mathcal HWW}/c_W^2 g_{\mathcal HZZ} = 1$, which also serves as a measure of the departure from custodial symmetry~\cite{CMS-PAS-HIG-14-009,ATLAS-CONF-2014-009}.

However, a priori, we are not limited to doublet representations and it is interesting to consider whether representations larger than a doublet of $SU(2)_L$ can significantly contribute to EWSB while satisfying $\rho \approx 1$, as well as LHC and Tevatron experimental constraints.~Care must be taken in these cases since the neutral components of non-doublet Higgs bosons will in general contribute to deviations from $\rho = 1$ at tree-level, when they acquire a VEV.~This, generically, imposes severe constraints on the size of the VEVs in order to obtain a value for $\rho$ in agreement with LEP measurements~\cite{Beringer:1900zz} and makes the corresponding models very fine-tuned and unappealing.

A well-known scenario which is free of this problem was proposed almost thirty years ago by Georgi and Machacek (GM)~\cite{Georgi:1985nv} and subsequently studied in detail in a number of early papers~\cite{Chanowitz:1985ug,Gunion:1989ci,Gunion:1990dt} as well as more recently in~\cite{Godfrey:2010qb,Englert:2013zpa,Aoki:2007ah,Carmi:2012in,Chiang:2012cn,Kanemura:2013mc,Chang:2012gn,Chiang:2013bqa,Killick:2013mya,Belanger:2013xza,Englert:2013wga,Chiang:2014bia}.~This model possesses the simplest extra non-doublet~\footnote{One could consider more exotic \emph{single} irreducible representations of $SU(2)_L\times U(1)_Y$ larger than doublets which also satisfy the condition necessary for $\rho = 1$ even if the VEVs are misaligned~\cite{Tsao:1980em,Gunion:1989we,Low:2010jp,Hisano:2013sn,Kanemura:2013mc,Alvarado:2014jva}, but we will not do so here.} representation of $SU(2)_L\otimes U(1)_Y$ which can participate \emph{non-negligibly} to EWSB, while remaining consistent with $\rho \approx 1$ thanks to the custodial symmetry imposed at the weak scale.

However, the GM model itself is not free of problems.~For one, the hierarchy problem of the SM is aggravated by virtue of the presence of extra light scalars.~Additionally, there are issues with maintaining custodial symmetry once radiative effects are considered~\cite{Gunion:1990dt}.~As it was suggested, a natural solution to these problems is to construct a supersymmetric version of the GM model as formulated recently in~\cite{Cort:2013foa}.~This model includes the same superfield content of the MSSM plus three $SU(2)_L$ Higgs triplet superfields with hypercharges $Y=0,\pm 1$.~They are arranged in such a way that all Higgs self interactions preserve a tree-level global $SU(2)_L\otimes SU(2)_R$ symmetry at some energy scale, $\mathcal M$.~We refer to this model as the supersymmetric custodial triplet model (SCTM). 

In this Letter we extend the initial tree-level study of the SCTM~\cite{Cort:2013foa}, which focused on the region $\mathcal M \sim v$, by performing a renormalization group evolution analysis and considering a large range of scales $\mathcal M$.~We present the main phenomenological features of the model at the tree-level improved by the renormalization group equations.~In particular, we show how the SCTM can be consistent with electroweak precision measurements even if the scale at which the $SU(2)_L\otimes SU(2)_R$ symmetry holds is in the multi-hundred TeV range and the Higgs triplets contribute sizably to EWSB.~As a by product of supersymmetry (SUSY), we also show that this can be made consistent with a 125 GeV Higgs mass, with top squarks generically lighter and with smaller mixing than those needed in the MSSM.

The paper is organized as follows.~In~\sref{2}, we briefly review the GM model and discuss its issues with naturalness and custodial symmetry at the electroweak scale.~In~\sref{3}, we introduce the field content of the SCTM and discuss how it addresses these issues.~In~\sref{rho}, we examine the electroweak vacuum and show that $\rho \approx 1$ can be accommodated even without custodial symmetry at the electroweak scale and with sizable contributions from the Higgs triplets to EWSB.~In~\sref{mh} we discuss how the observed Higgs mass of $\sim$ 125 GeV can be easily reproduced via sizable tree-level contributions from additional $F$-terms and one-loop corrections which are generically larger than those arising in the MSSM.~In~\sref{lhc}, we discuss the `smoking guns' of the SCTM at the LHC and in particular the departure from the universal condition of the Higgs couplings to $Z$ and $W$ bosons.~Finally in~\sref{disc} we give our conclusions and outlook.~More details and results, including one-loop corrections, will be published in a more extensive study~\cite{preparation}.

\section{Custodial Symmetry in\\ the GM Model}\label{sec:2}
In the GM model, two $SU(2)_L$ triplets scalars are added to the SM in such a way that the Higgs potential preserves a global $SU(2)_L\otimes SU(2)_R$ symmetry which is broken to the vector custodial~\footnote{Often `custodial' refers to both the global $SU(2)_L\otimes SU(2)_R$ symmetry and $SU(2)_V$ subgroup interchangeably.~Here we will explicitly distinguish between them since the custodial symmetry is a symmetry of the gauge boson mass matrix and thus it is only well defined at the weak scale, while the global $SU(2)_L\otimes SU(2)_R$ can in principle be imposed at any scale.} subgroup $SU(2)_V$ after EWSB, predicting $\rho = 1$ at the tree-level~\cite{Georgi:1985nv}.~More specifically, on top of the SM Higgs doublet $H=(H^+,H^0)^T$, one real $SU(2)_L$ triplet scalar with hypercharge $Y=0$, $\phi=(\phi^+,\phi^0,\phi^-)^T$, and one complex triplet scalar with $Y = 1$,  $\chi=(\chi^{++},\chi^+,\chi^0)^T$, are added.~In terms of representations of $SU(2)_L\otimes SU(2)_R$ we have,
\begin{equation}
H=\left(\begin{matrix}
H^{0*} & H^+\\
H^-&H^0
\end{matrix}  
\right),\quad
\Delta=\left(\begin{matrix}
\chi^{0*} & \phi^+& \chi^{++}\\
\chi^-&\phi^0&\chi^+\\
\chi^{--}&\phi^-&\chi^0 
\end{matrix}  
\right),
\end{equation}
transforming as $({\bf 2,2})$ and $({\bf 3,3})$, respectively.

If EWSB proceeds such that $v_H\equiv \langle H^0\rangle$, $v_\phi\equiv\langle \phi^0\rangle=v_\chi\equiv\langle \chi^0\rangle$,~i.e.~the triplet VEVs are aligned, then $SU(2)_L\otimes SU(2)_R$ will be broken to the custodial subgroup $SU(2)_V$, which ensures that the $\rho$ parameter is equal to one at tree-level as in the SM.~This can be explicitly seen by computing the deviation from $\rho_{tree} = 1$ when the triplet VEVs have a generic configuration,
\bea
 \rho_{tree} -1 \equiv \Delta\rho = \frac{2(v_\phi^2-v_\chi^2)}{v_H^2+4v_\chi^2} .
\label{eq:rhoGM}
\eea
Thus, having custodial symmetry, which requires $v_\phi = v_\chi$, is equivalent to the condition $\rho = 1$ at tree-level.~Moreover, by imposing custodial symmetry, one easily finds the tree-level relation for the Higgs couplings to gauge bosons,
\bea
\frac{g_{\mathcal H WW}}{c_W^2g_{\mathcal HZZ}}=1 ,
\label{eq:universality}
\eea
where $\mathcal H$ is either of the custodial singlets which contribute to EWSB.~The tree-level universality behavior of~\eref{universality} is implied by the condition $\rho_{tree}=1$, and thus it is extremely constrained by electroweak precision data. 

However, it is important to note that only the tree-level Higgs sector is invariant under the $SU(2)_L\otimes SU(2)_R$ global symmetry.~The Yukawa and hypercharge interactions lead to an explicit breaking of this symmetry by radiative corrections.~Thus, even if the Higgs sector of the theory is $SU(2)_L\otimes SU(2)_R$ invariant at one particular scale, in general it will be driven, by the renormalization group equation (RGE) evolution of the couplings and mass parameters, to a point which violates this global symmetry.

In the GM model, this implies that, if the scale at which $SU(2)_L\otimes SU(2)_R$ holds (which we call $\mathcal M$) is far above the electroweak scale, RGE evolution will typically lead to large deviations from $\rho_{tree} = 1$ at the electroweak scale, in conflict with experiments.~Thus in the GM model, one is forced to impose the scale $\mathcal M$, which is \emph{a priori} unrelated to $v$, to be close to the electroweak scale.~The particular choice of the scale $\mathcal M$ will also greatly affect the phenomenology of the model~\cite{Gunion:1989ci,Gunion:1990dt}.

We also emphasize that there should be new dynamics at the scale $\mathcal{M}$ where the $SU(2)_L\otimes SU(2)_R$ symmetry is imposed.~Otherwise, this $SU(2)_L\otimes SU(2)_R$ symmetric point is simply an arbitrary point in the RGE evolution which `accidentally emerges' via running from some $SU(2)_L\otimes SU(2)_R$ violating point at higher energies, a scenario we find unappealing.~In other words, to avoid relying on this accidental emergence of the global $SU(2)_L\otimes SU(2)_R$, the scale $\mathcal{M}$ should also be taken as the cutoff of the theory.~In the GM model this implies a cutoff at or around the electroweak scale, or the introduction of new dynamics, or degrees of freedom,~\emph{beyond} those found in the GM model, such as a strongly coupled sector as originally proposed in the GM model~\cite{Georgi:1985nv}.~These problems can be seen as an indication that the GM model should be embedded in a larger theory which would presumably resolve these issues.

\section{Custodial Symmetry in \\ the Supersymmetric Custodial\\ Higgs Triplet Model}\label{sec:3}
In this section, we briefly review the SCTM field content and discuss how the model alleviates the various issues of the GM model.~In addition to the two MSSM Higgs doublets $H_1$ (coupled to down quarks and leptons) and $H_2$ (coupled to up quarks), we add three complex triplets $\Sigma_0=(\phi^+,\phi^0,\phi^-)^T$, $\Sigma_+=(\psi^{++},\psi^+,\psi^0)^T$, and $\Sigma_{-}=(\chi^0,\chi^-,\chi^{--})^T$, with hypercharge $Y=0,+1,-1$, respectively, corresponding to the two triplets $\phi$ and $\chi$ of the GM model.~After defining the $H\equiv ({\bf 2,2}) \equiv(H_1,H_2)$ and $\Delta \equiv ({\bf 3,3}) \equiv(\Sigma_-,\Sigma_0,\Sigma_+)$ representations of $SU(2)_L\otimes SU(2)_R$, the bi-doublets and bi-triplets decompose under $SU(2)_V$ as $({\bf 2,2})={\bf 1\oplus 3}$ and $({\bf 3,3})={\bf 1\oplus 3\oplus 5}$ which provides a classification of mass eigenstates in the custodial theory after EWSB~\cite{Cort:2013foa}.

When the neutral components of the doublet and triplet fields develop VEVs $v_1=\langle H_1^0\rangle,\,v_2=\langle H_2^0\rangle,\,v_\phi=\langle\phi^0\rangle,\,v_\psi=\langle \psi^0\rangle,\,v_\chi=\langle \chi^0\rangle$, the deviation from $\rho_{tree} = 1$ is given by,
\bea
\Delta\rho=\frac{2( 2v_\phi^2-v_\psi^2-v_\chi^2)}{v_1^2+v_2^2+4(v_\chi^2+v_\psi^2)} .
\label{eq:rhoSCTM}
\eea
As it can be seen, $\Delta\rho=0$ if custodial symmetry is preserved at the minimum of the theory which requires $v_1=v_2$ and $v_\phi=v_\psi=v_\chi$.~However, unlike in the GM model, custodial symmetry is no longer a necessary (although certainly sufficient) condition for $\rho_{tree}=1$, that is also satisfied along the non-custodial direction $2v_\phi^2=v_\psi^2+v_\chi^2$.~This `extra direction' for the VEVs is a consequence of supersymmetry where the $Y = 1$ and $Y = -1$ triplets are separate fields with, in general, distinct VEVs~\footnote{A similar situation happens in the MSSM where the SM custodial symmetry is broken if $v_1\neq v_2$, but in this case the breaking enters only at one loop and thus $\rho_{tree} =1$~\cite{Drees:1990dx}.} in contrast to the GM model where they make up one complex field with hypercharge $Y = 1$.~As a consequence, the universal relation for the Higgs couplings in~\eref{universality} is no longer implied by the experimentally measured value of $\rho \approx 1$.~Furthermore, as we will see in next section, this additional direction allows us to have the scale $\mathcal M$ at which the $SU(2)_L\otimes SU(2)_R$ symmetry is imposed to be much higher than the electroweak scale.

We also point out that a natural choice for the scale $\mathcal M$ is the messenger scale $M$~\footnote{Supersymmetry is assumed to be broken in a hidden sector, where an $F$ (or $D$) term acquires a VEV, and communicated to the observable sector by messenger fields of mass $M$ where $F\ll M^2$.~The mass of the superpartners $m_{\tilde f}$ is thus proportional to $F/M$ with a coefficient which depends on the dynamics of the transmission (e.g.~tree-level versus loop-level).~After integrating out the messenger fields the effective theory is a supersymmetric one with soft breaking masses $m_{\tilde f}$ and cutoff at the scale $M$.~As a consequence the inequality $m_{\tilde f}\ll M$ holds.}.~Thus, unlike in GM model, this allows $\mathcal M$ to now be associated with a physical scale which, once known, can be used to predict the value of $\rho_{tree}$ at the electroweak scale through RGE evolution.~Conversely, a measurement of $\rho$ now gives a constraint on the scale of SUSY breaking.
Taking $\mathcal{M}$ to be below the messenger scale $M$ reintroduces the accidental emergence problem of the global $SU(2)_L\otimes SU(2)_R$ described in the previous section.~In principle, one could take $\mathcal{M}$ to be above the messenger scale, but this would require assumptions about the SUSY breaking mechanism.~Since we are not attempting to explicitly construct such a mechanism, we simply take $\mathcal{M}$ to be at the messenger scale, and assume that the mechanism which breaks SUSY also generates the $SU(2)_L\otimes SU(2)_R$ invariant Higgs sector.~Therefore, we are making the assumption that the messenger sector, which transmits supersymmetry breaking to the observable sector, exhibits the $SU(2)_L\otimes SU(2)_R$ invariance and then proceeds through effective operators as,
\begin{equation}
\int d^4\theta \frac{X^\dagger X}{\mathcal M^2}Y^\dagger Y,\quad Y=H,\Delta, Q, L, U^c,D^c,E^c,
\end{equation}
where~$X=\theta^2 F$ is the spurion superfield responsible for supersymmetry breaking.

\section{The Electroweak Vacuum and tree-level $\rho$ Parameter}\label{sec:rho}

We now examine the Higgs potential and the electroweak vacuum of the SCTM to show how sizable values of triplet VEVs and high scales $\mathcal M$ are allowed by constraints on the $\rho$ parameter.~Because of the explicit breaking by the hypercharge and Yukawa~\footnote{We implicitly assume global lepton number conservation so that the supersymmetric $SU(2)_L\otimes SU(2)_R$ violating operator $\Sigma_+ LL$ is forbidden, but in principle it can be included as part of a model to generate neutrino masses~\cite{FileviezPerez:2012ab,FileviezPerez:2012gg}.~We also do not consider possible Dirac gaugino mass terms of the form $m_D \widetilde{\Sigma}_0^a \widetilde{W}^a$ which would violate the global $SU(2)_L\otimes SU(2)_R$.These terms could appear from $D$-term supersymmetry breaking corresponding to a hidden $U(1)'$ whose chiral density breaks supersymmetry as $W'_\alpha=\theta_\alpha D$ and the effective operator $(1/M)\,\int d^2\theta W'_\alpha W_a^\alpha \Sigma_0^a$ yields a Dirac gaugino mass.
We just assume that the UV completion of the SCTM can explain its absence.} interactions, the superpotential is in general not $SU(2)_L\otimes SU(2)_R$ invariant.~In terms of the neutral components of the Higgs doublets $(H_1^0,H_2^0)$ and triplets $(\psi^0,\phi^0,\chi^0)$ it is given by,
\bea
W^0&=&
\lambda_a H_1^0 \psi^0 H_1^0 + \lambda_b H_2^0 \chi^0 H_2^0 + \lambda_c H_1^0 \phi^0 H_2^0 \nonumber \\
&+& \lambda_3 \psi^0 \phi^0\chi^0 - \mu H_1^0 H_2^0 + \frac{\mu_a}{2} (\phi^0)^2 + \mu_b\psi^0\chi^0 .
\label{eq:superpotential}
\eea
The scalar potential is then $V=V_F+V_D+V_{soft}$ where $V_F=\sum_X |\partial W^0/\partial X|^2$ and  $X=H_1^0,H_2^0,\psi^0,\phi^0,\chi^0$ while the $D$-terms are given by,
\bea
\label{eq:vD}
V_D=\frac{g_2^2+g_1^2}{8}(|H_1^0|^2-|H_2^0|^2+2 |\chi^0|^2-2|\psi^0|^2)^2,
\eea
where $g_2$ and $g_1$ are the $SU(2)$ and $U(1)_Y$ couplings, respectively.~Finally, the soft SUSY breaking terms are given by,
\bea
\label{eq:vsoft}
V_{soft} &=& 
m_{H_2}^2|H_2^0|^2+m_{H_1}^2|H_1^0|^2 \nonumber \\
&+& m_{\Sigma_{-1}}^2|\chi^0|^2 + m_{\Sigma_0}^2|\phi^0|^2 + m_{\Sigma_1}^2|\psi^0|^2  \\
&+& \Big( A_a H_1^0 \psi^0 H_1^0+A_b H_2^0 \chi^0 H_2^0 \nonumber \\
&+& A_c H_1^0 \phi^0 H_2^0 + A_3 \psi^0\phi^0\chi^0 \nonumber \\
&-& m_3^2 H_1^0 H_2^0 + B_a (\phi^0)^2/2 + B_b \psi^0\chi^0 + H.c.\Big) . \nonumber
\eea
The global $SU(2)_L\otimes SU(2)_R$ invariance of the Higgs sector translates into the following boundary conditions at the scale $\mathcal{Q}=\mathcal M$,
\begin{align}\label{eq:bdcns}
\lambda_a&=\lambda_b=\lambda_c\equiv \lambda,\quad
\mu_a=\mu_b\equiv \mu_\Delta\nonumber\\
m_{H_1}&=m_{H_2}\equiv m_H,\quad
m_{\Sigma_0}=m_{\Sigma_1}=m_{\Sigma_{-1}}\equiv m_\Delta\nonumber\\
A_a&=A_b=A_c\equiv A_\lambda,\quad
B_a=B_b\equiv B_\Delta .
\end{align}
In the limits $|B_\Delta| \to \infty$ and $m_3^2 \to \infty$ and when $\mathcal{M} \sim v$ it is possible to recover the scalar spectrum found in the GM model~\cite{Cort:2013foa}.~However, as we discuss below, since we generically have $\mathcal{M} > v$, the scalar spectrum of the SCTM will typically look quite different from the one found in the GM model. 

Once the boundary conditions in~\eref{bdcns} are imposed, we then run from $ \mathcal{Q}=\mathcal{M}$ down to the scale $\mathcal{Q}_{EW}\equiv m_t$,~where $m_t$ is the top mass~\footnote{There are in principle threshold effects which should be accounted for in the RG running from $\mathcal{M}$ to the electroweak scale.~However, unless $\mathcal{M} \sim$ TeV where our new spectrum lies, these effects are expected to be small~\cite{Drees:1990dx} and are therefore neglected.~Nevertheless, a precise analysis of the region $\mathcal{M} \sim$ TeV should include these corrections.}, and solve the equations of minimum (EOM) for the scalar potential corresponding to the five neutral field directions $(H_1^0,H_2^0,\psi^0,\phi^0,\chi^0)$.~We can then parametrize the minimum by two VEVs $(v_H,v_\Delta)$ and three angles $(\beta,\theta_1,\theta_0)$ as,
\bea
v_1&=& \sqrt{2}\cos\beta v_H,~~v_2=\sqrt{2}\sin\beta v_H, \nonumber\\
v_\psi &=& 2\cos\theta_1\cos\theta_0 v_\Delta,~~v_\chi=2\sin\theta_1\cos\theta_0 v_\Delta,\nonumber\\
v_\phi &=& \sqrt{2} \sin\theta_0 v_\Delta .
\label{eq:vacio}
\eea
With this parametrization, custodial symmetry is controlled by the three angles $(\beta,\theta_0,\theta_1)$ where in the custodial limit, $\tan\beta=\tan\theta_0=\tan\theta_1=1$.~On the other hand looking at deviations from $\rho_{tree} = 1$ we find that the dependence on $\theta_1$ and $\beta$ cancels out leaving only a dependence on $\theta_0$ given by,
\beq\label{eq:deltaRhoAndTheta0}
\Delta\rho=-4\frac{\cos 2\theta_0 v_\Delta^2}{v_H^2+8\cos^2\theta_0v_\Delta^2} .
\eeq

For our analysis, given the boundary conditions at the scale $\mathcal{M}$, we will consider $\mathcal M$ and $v_\Delta$ as free parameters.~Then the value of $v_H$ is determined by the experimental measurements of the $W$ mass, leading to the constraint on the EW scale $v^2 = 2 v_H^2 + 8 v_\Delta^2$~\cite{Cort:2013foa}, where $v = 174$ GeV.~As the parameters $m_3^2$ and $B_{a,b}$ have their RGEs decoupled from the rest of the parameters, we can consistently fix two parameters $m_3^2$ and $B_+\equiv B_a+B_b$ from their respective EOMs.~The other three EOM (including the one for $B_-\equiv B_a-B_b$), which vanish identically in the custodial limit, self-consistently determine the values of the custodial breaking angles  $(\tan\beta,\tan\theta_0,\tan\theta_1)$, which are therefore a prediction of the EOMs for given values of $v_\Delta$ and $\mathcal{M}$.

For illustrative purposes, we will consider an example parameter point by fixing the following parameters at the high scale $\mathcal M$ (as in~\cite{Cort:2013foa}),
\bea
\lambda_3&=&-0.35,~\mu= \mu_\Delta=250\,\text{GeV}, \nonumber \\
A_\lambda &=& A_3=A_t=A_b \equiv A_0 = 0,  \nonumber \\
~m_H &=& m_\Delta=1000\,\text{GeV},\,M_1=M_2=M_3\equiv m_{1/2},\nonumber \\
~m_Q&=&m_{U^c}=m_{D^c}\equiv m_0=500\,\text{GeV} .
\label{eq:valores}
\eea
Our results will be shown for different values of $m_{1/2}$: (1, 1.1, 1.2, 1.3) TeV~\footnote{Since the values for the squark masses and for the gluino mass $M_3$ increase as we run to lower scales, we find that our benchmark point leads to a spectrum that satisfies current direct search constraints from the LHC searches.~However, a detailed analysis of the LHC phenomenology is beyond the scope of this Letter.}.~As we discuss more in detail in the next section, the parameter $\lambda$ is fixed by the condition that the Higgs field dominantly responsible for EWSB $\mathcal H$ has a mass of $\sim 125$ GeV.
\begin{figure}[htb]
\begin{center}
\includegraphics[width=85mm]{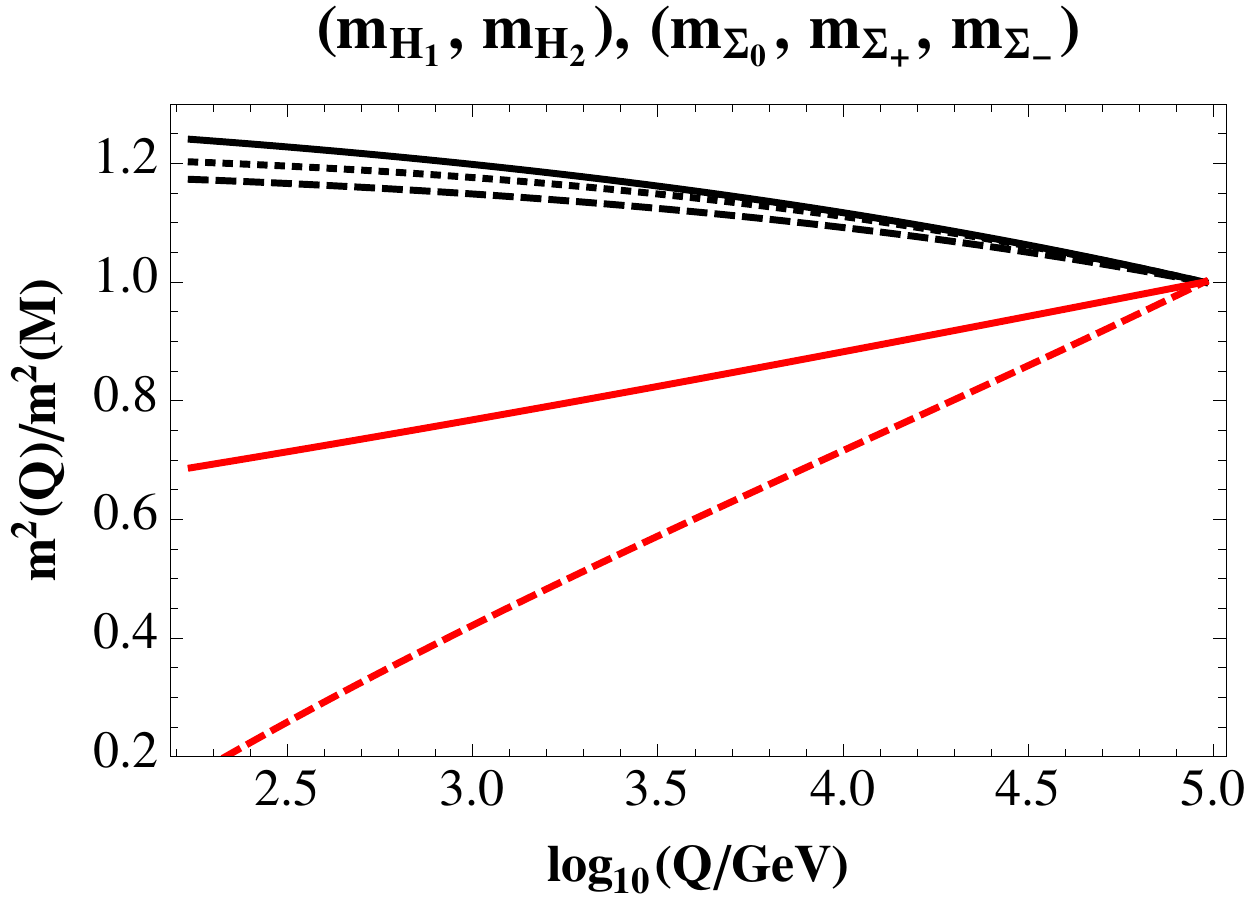}
\end{center}
\caption{\it 
Running of  $(m_{H_1}^2,m_{H_2}^2)$ (red lines from top to bottom) and $(m_{\Sigma_0}^2,m_{\Sigma_+}^2,m_{\Sigma_-}^2)$ (black lines from bottom to top), normalized to their values at the scale $\mathcal M=10^5$ GeV for $m_{1/2}=1.2$ TeV and $v_\Delta = 20$~GeV.}
\label{fig:running}
\end{figure}

We show in~\fref{running} the results of the RGE running parameters $(m_{H_1}^2,m_{H_2}^2)$ (red lines from top to bottom) and $(m_{\Sigma_0}^2,m_{\Sigma_+}^2,m_{\Sigma_-}^2)$ (black lines from bottom to top), normalized to their values at the scale $\mathcal M$ (chosen to be $10^5$ GeV), as functions of the RG scale $\mathcal Q$ ($<\mathcal M$) and for $v_\Delta = 20$~GeV,~$m_{1/2}=1.2$ TeV.~The dispersion in $(m_{H_1}^2,m_{H_2}^2)$, which is responsible for generating $\tan\beta\neq 1$ at $\mathcal Q_{EW}$, is much larger than the dispersion in the sector $(m_{\Sigma_0}^2,m_{\Sigma_+}^2,m_{\Sigma_-}^2)$, that is responsible for the departure of $\tan\theta_{0}$ and $\tan\theta_{1}$ from their custodial values.~This is because the largest contribution to the doublet splitting comes from the custodial breaking by the top and bottom Yukawa sectors to which the doublet couples at tree-level.~The splitting in the triplet sector is instead mainly driven by the hypercharge interactions since triplets do not couple to the top and bottom sectors at tree-level.~Thus the splitting in the triplet mass parameters due to the top and bottom Yukawa interactions is only a higher order effect.~This gives in general $|\tan\theta_0-1|,|\tan\theta_1-1|<|\tan\beta-1|$.~Since $\Delta\rho$ only depends on $\tan\theta_0$ (see~\eref{deltaRhoAndTheta0}) we expect deviations from $\rho_{tree} = 1$ to be small as well.

These features can be seen by examining~\fref{T} and~\fref{theta0}.~In~\fref{T} we show the regions allowed at the $95\%$ C.L.~by the experimental value of the $T$ parameter ($\Delta\rho=\alpha T$), corresponding to the fit value $T=0.07\pm 0.08$~\cite{Beringer:1900zz}.~We show results for various values of the common gaugino mass $m_{1/2}=$ 1 (black lines), 1.1 (blue lines), 1.2 (red lines) and 1.3 (orange lines) TeV, at the scale $\mathcal M$.~The allowed region is inside the corresponding solid lines with the dashed lines indicating the $T=0$ contour. One could interpret the funnel regions that appear for large $v_\Delta$ values as a fine tuning of the scale $\mathcal{M}$. However in the absence of a precise theory of supersymmetry breaking one could also interpret these regions as a precise prediction of the scale $\mathcal{M}$ which should be provided by the underlying supersymmetry breaking sector.~We also show the low $SU(2)_L\otimes SU(2)_R$ scale $\mathcal M$ region in~\fref{T} only for illustrative purposes to demonstrate that, as in the GM model, the parameter space for $v_\Delta$ opens up considerably as $\mathcal{M}\rightarrow v$.~A proper treatment of this region should also include threshold corrections in the RG running.~Furthermore, one must ensure that the physical particle masses are below $\mathcal{M}$ which is a consistency condition since, as discussed above, $\mathcal{M}$ serves as the cutoff for the theory. 

We see at this point that the extra freedom (the VEV direction $2v_\phi^2=v_\chi^2+v_\psi^2$) in the SCTM, with respect to the non-supersymmetric GM model, comes into play allowing for $T = 0$ contours (along dashed lines) throughout the parameter space.~In fact, generically the three VEVs $v_\phi,v_\psi,v_\chi$ are not equal along the $T = 0$ contours.~The new direction allows for scales well above $\sim 100$ TeV and sizable triplet VEVs to be comfortably within the allowed region.~These $T = 0$ contours will shift slightly after including the sub-dominant one-loop corrections, using the RGE improved Lagrangian, but we do not investigate this issue here.
\begin{figure}[htb!]
\begin{center}
\includegraphics[width=85mm]{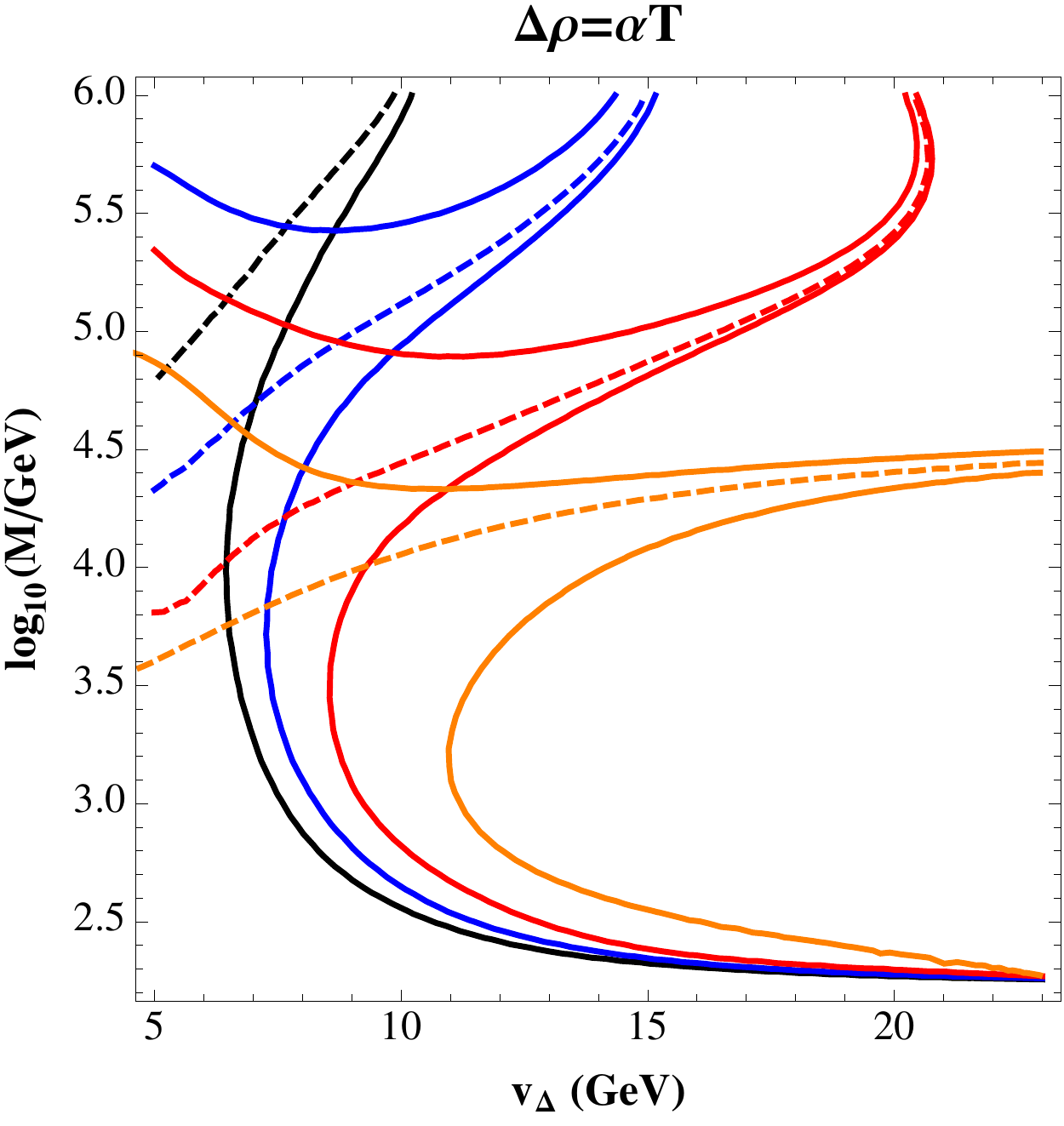}
\end{center}
\caption{\it Regions allowed by the $T$ parameter as a function of $\mathcal M$ and $v_\Delta$.~The region between the solid lines corresponds to the allowed 95\% CL interval, having fixed the parameters as in~\eref{valores} and for $m_{1/2}=$1 (solid black lines), 1.1 (solid blue lines), 1.2 (solid red lines) and 1.3 (solid orange lines) TeV at the scale $\mathcal M$.~The corresponding dashed lines are for $T=0$.}
\label{fig:T}
\end{figure}

In~\fref{theta0} we show contours of $\tan\beta$ (blue dashed), $\tan\theta_0$ (black solid), and $\tan\theta_1$ (dark green dotted).~The shaded region is the one allowed by the $T$ parameter at the 95\% CL for $m_{1/2} = 1.2$~TeV.~As expected from~\eref{deltaRhoAndTheta0}, in the region allowed by the $\rho$ parameter, deviations from $\tan\theta_0=1$ are very small.~Furthermore, as anticipated from the results of the running in~\fref{running}, the violation of custodial symmetry is much larger in $\tan\beta$, which can have values as large as $\tan\beta\gtrsim 2$, than for the parameters $\tan\theta_0$ and $\tan\theta_1$ which depart from their custodial values only by a few percent.~We note the presence of a `crossover' point where the triplet VEVs are aligned $\tan\theta_0=\tan\theta_1=1$, as found in the GM model.~This limit is not equivalent to the GM model, however, since the scale $\mathcal{M}$ is still much greater than the electroweak scale.~After RGE running this will lead to a significantly different scalar spectrum at the electroweak scale from the one found in the GM model. 
\begin{figure}[htb!]
\begin{center}
\includegraphics[width=85mm]{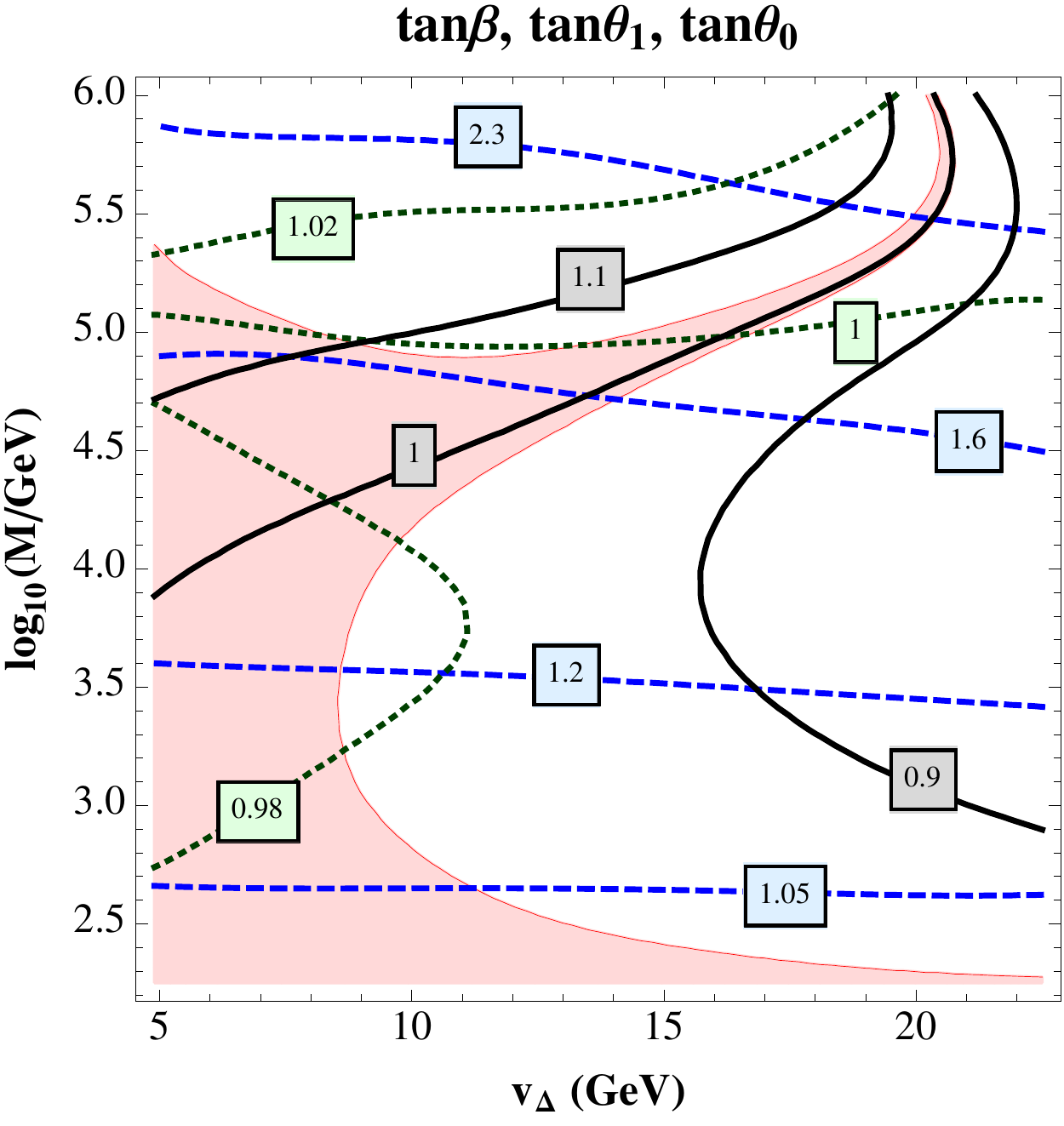}
\end{center}
\caption{\it Contours of $\tan\beta$ (blue dashed), $\tan\theta_0$ (black solid), and $\tan\theta_1$ (dark green dotted) are shown, having fixed the parameters as in~\eref{valores} and for $m_{1/2}=1.2$ TeV.~Shaded pink region is allowed at 95 \% CL by the $T$ parameter. }
\label{fig:theta0}
\end{figure}

We emphasize that the SCTM is free of generic issues found in supersymmetric models with only one Higgs triplet, which in general acquires a VEV that must be tuned to be very small~($\sim$3 GeV at 95\% CL, Ref.~\cite{Beringer:1900zz}, for our normalization choice, $v=174$ GeV) in order to satisfy electroweak precision data (see for example~\cite{Delgado:2012sm,Delgado:2013zfa}).~In contrast, in the SCTM, one can obtain triplet VEVs as large as $\sim 25$ GeV (possibly larger if $\mathcal M \sim v$).~Although $25$ GeV does not appear large, the actual contribution to the electroweak symmetry breaking is much larger, as can be seen from the condition $v^2 = 2 v_H^2 + 8 v_\Delta^2$.~For $v_\Delta = 25$ GeV this gives a $\sim 15\%$ contribution to EWSB which is significantly larger than the $\mathcal{O}(0.1\%)$~\cite{Beringer:1900zz} contribution allowed by the $\rho$ parameter in conventional triplet extended SUSY models.

Finally, we also point out that $v_\Delta$ is bounded from above by the condition of perturbativity of the top Yukawa coupling.~Since the top sector obtains its mass at tree-level only from the $SU(2)_L$ doublet VEV $v_2$, large values of $v_\Delta$ necessitate large top Yukawa couplings at the electroweak scale~\cite{Cort:2013foa} in order to reproduce the observed top mass.~One can see this by writing the top Yukawa coupling in terms of $v_\Delta$ as,
\bea\label{eq:topY}
h_t = \frac{m_t}{v_2}=\frac{m_t}{\sin\beta\sqrt{v^2 - 8 v_\Delta^2}},
\eea
which leads to the absolute constraint $v > 2 \sqrt 2 v_\Delta \Rightarrow v_\Delta \lesssim 62$ GeV.~Furthermore, if we demand $h_t \lesssim 4\pi$ at the scale $\mathcal M$, then it is typically difficult to get values for $v_\Delta$ much larger than $\sim 30$ GeV if we want to have a scale as high as $\mathcal M=\mathcal O(100\,\rm{TeV})$, since $h_t$ increases when run up to higher energies.

\section{The Higgs Boson Mass}\label{sec:mh}

Apart from electroweak data, the model needs to be contrasted with LHC data and in particular with measurements of the Higgs properties at the LHC.~We postpone a systematic analysis of Higgs and LHC observables and instead focus on a subset of observables which reflect the essential features of the model beginning with the experimentally measured Higgs mass.

The observed $\mathcal H\to ZZ$ and $\mathcal H\to WW$ decay rates~\cite{CMS-PAS-HIG-14-009,ATLAS-CONF-2014-009} suggest the Higgs giving the dominant contribution to EWSB is the custodial singlet primarily coming from the $({\bf 2}, {\bf 2})$ electroweak doublet and we will assume this to be true in what follows.~In the SCTM this Higgs is generally the lightest scalar in the spectrum and, in particular, it is the lighter of the two custodial singlets which trigger EWSB~\cite{Cort:2013foa}.~This is in contrast to the typically studied GM model, which has an additional $Z_2$ symmetry in the scalar potential~\cite{Gunion:1990dt,Englert:2013zpa,Englert:2013wga}, where the lightest scalar is the custodial singlet which has the \emph{least} to do with EWSB.~On the other hand if one considers the most general scalar potential allowed in the GM model, which also possesses a decoupling limit~\cite{Hartling:2014zca}, then the custodial singlet driving EWSB can be the lightest scalar.~This allows for the GM model to be recovered as a limit of the SCTM when $\mathcal M \sim v$.

Additionally, the SCTM possesses a feature shared with conventional triplet extended MSSM scenarios~\cite{Espinosa:1991gr,Espinosa:1991wt,Espinosa:1992hp,Kane:1992kq,Quiros:1998bz,Basak:2012bd,Bandyopadhyay:2013lca,Bandyopadhyay:2014tha} in that the SM-like Higgs mass can be pushed up by additional $F$-terms, and therefore does not have to rely heavily on large radiative corrections, as in the MSSM.~The $F$-terms are generated through the quartic couplings $\lambda_{a,b,c}$ and lead to a contribution at tree-level to the Higgs mass which, in the decoupling limit, is proportional  to~$4\lambda_a^2\cos^4\beta+4\lambda_b^2\sin^4\beta + \lambda_c^2\sin^2 2\beta$.~Furthermore, since radiative corrections to the squared Higgs mass coming from top squarks are $\propto h_t^4$, using~\eref{topY} we see that for $v_\Delta > 0$ they are enhanced with respect to the MSSM contribution.~Thus the SCTM allows in general for larger tree-level \emph{and} one-loop contributions to the Higgs mass than those that can be found in the MSSM.~Note also that in the custodial limit where $\tan\beta = 1$ there is no tree-level contribution from the doublet (or MSSM) sector to the Higgs mass. 

It is also important to ensure that the correct Higgs mass can be reproduced with perturbative values of $\lambda$.~To see this we show in~\fref{lambda} contour lines of $\lambda$ (defined at the high scale $\mathcal M$) reproducing the observed Higgs mass, including the stop loop corrections, in the $(v_\Delta,\mathcal M)$ plane for the benchmark point in~\eref{valores} and fixed $m_{1/2}=1.2$ TeV.~A Higgs mass of $\sim 125$~GeV can be obtained for messenger scales $\gtrsim 100$ TeV and triplet VEVs as large as $v_\Delta \sim 25$ GeV over a range of perturbative values for $\lambda$.~Taking as an example $\lambda = 0.5$, $v_\Delta \sim 25$ GeV, and $\mathcal M \sim 100$ TeV gives a tree-level contribution to the Higgs mass $\sim 100$ GeV which is larger than $m_Z$, the absolute upper bound on the tree-level contribution allowed in the MSSM.

Here we do not perform a general parameter space analysis, but comment that a number of competing effects lead to the features seen in~\fref{lambda}, both at tree-level through $\lambda$ and radiatively through enhanced stop corrections at large $v_\Delta$, or large RGE effects for high scales of $\mathcal M$.~In particular, smaller values of $\lambda$ are equired at large $\mathcal M$.~This might be at first surprising since $\lambda$ (or more precisely $\lambda_{a,b,c}$) runs to smaller values as we go down from $\mathcal M$ to $\mathcal Q_{EW}$ implying small tree-level contributions from the triplet sector.~However, as we increase $\mathcal M$ beyond $ \gtrsim 10^4$~GeV, the increasing values of $\tan\beta$ from $\tan\beta = 1$ (see~\fref{theta0}) lead to the `turning on' of the tree-level MSSM contribution allowing for smaller values of $\lambda$ to be consistent with the observed Higgs mass.
\begin{figure}[htb!]
\includegraphics[width=84mm,height=80mm]{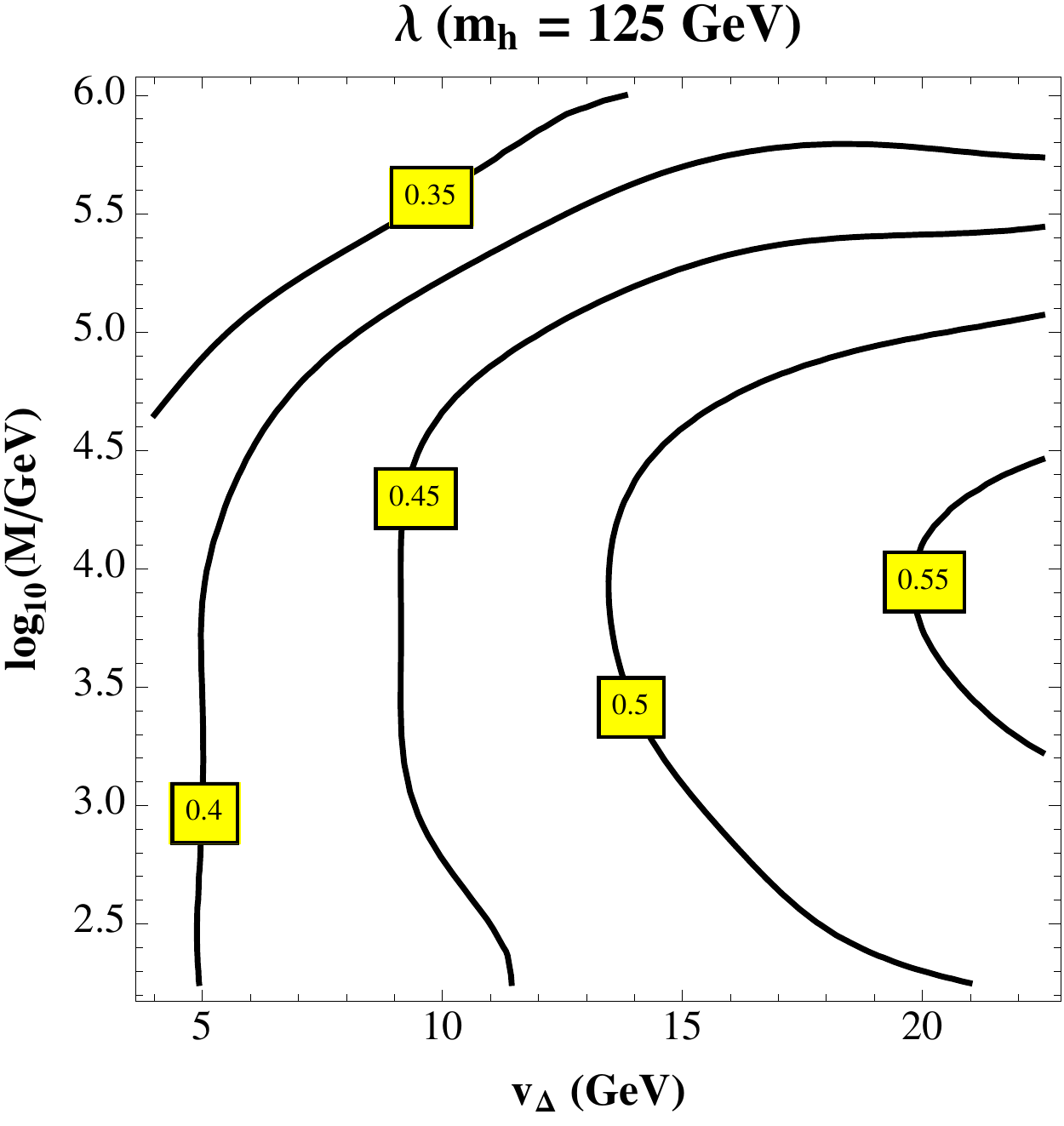}
\caption{\it 
Contours of $\lambda$, defined at the high scale $\mathcal M$, reproducing the observed value of the Higgs mass $\sim 125$~GeV for the $SU(2)_L\otimes SU(2)_R$ symmetric parameters in~\eref{valores} and $m_{1/2}=1.2$ TeV.}
\label{fig:lambda}
\end{figure}

We also examine whether light top squarks ($\lesssim 1$~TeV) together with small trilinear terms can be accommodated in the SCTM while still reproducing the observed Higgs mass, in contrast to the MSSM which requires large $A$-terms to avoid multi-TeV top squarks.~In~\fref{mstop} we show the allowed values for the physical lightest stop mass which reproduces a Higgs mass of $125.5\pm 1.0$~GeV, for the example parameter point, $\lambda=0.45$, $\mathcal M = 65$ TeV, $m_{1/2}=1.2$ TeV, $v_\Delta=10$ GeV and all other parameters fixed to the values in~\eref{valores}, except we now allow the soft and tri-linear mass parameters to be in the ranges $m_0 \in [500, 1000]$~GeV and $A_0 \in [-250, 500]$~GeV.~In the region allowed by the $\rho$ parameter (shaded pink in~\fref{mstop}) we see top squarks as light as $\sim 950$~GeV can produce the correct Higgs mass for modest values of the trilinear couplings at the electroweak scale $X_t\equiv A_t-\mu/\tan\beta\sim -750$ GeV.~These numbers should be compared to the MSSM prediction where for trilinear terms $\sim 1$~TeV,~and $\tan\beta \sim 20$, the top squarks should be heavier than $\sim 6$~TeV~\cite{Draper:2011aa,Feng:2013tvd,Delgado:2013gza,Draper:2013oza} showing that the SCTM indeed helps to alleviate the MSSM fine-tuning problem (see also~\cite{Agashe:2011ia}).
%
\begin{figure}[htb!]
\begin{center}
\Large{{\bf~~~~~~~~~~~~~~~~~~~$\rho\pm\delta\rho$, $m_h\pm\delta m_h$~~~~~~~~~~~}}\normalsize\\
\includegraphics[width=85mm]{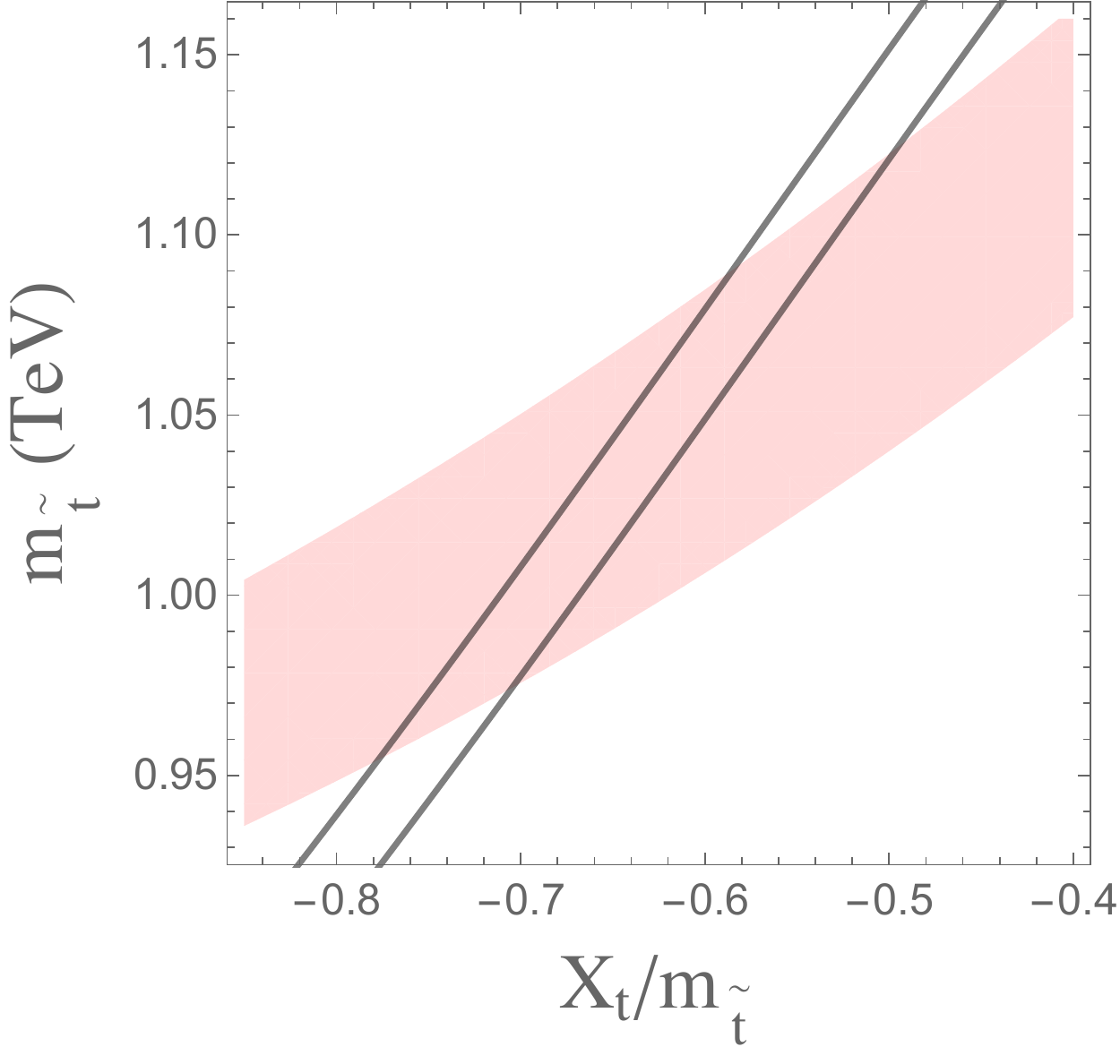}
\end{center}
\caption{\it The solid black lines represent the region producing a Higgs mass of $125.5\pm 1.0$~GeV in the $X_t/m_{\tilde t}-m_{\tilde t}$ plane, where $m_{\tilde t}$ is the physical mass of the lightest stop and $X_t\equiv A_t-\mu/\tan\beta$.~The shaded pink band is the region allowed by constraints on the $\rho$ parameter.~We have fixed the parameters $\lambda=0.45$, $\mathcal M = 65$ TeV, $m_{1/2}=1.2$ TeV, $v_\Delta=10$ GeV while the rest are given in~\eref{valores}, except we now allow $m_0 \in [500, 1000]$~GeV and $A_0 \in [-250, 500]$.~We do not explicitly show the region favored by the MSSM since it arises only at much heavier stop masses ($m_{\tilde t}\gtrsim 6$~TeV~\cite{Draper:2011aa,Feng:2013tvd,Delgado:2013gza,Draper:2013oza}).}
\label{fig:mstop}
\end{figure}

\section{Smoking Guns at LHC}\label{sec:lhc}

The next observables we consider, and potential smoking guns of the model at the LHC, are the normalized couplings of the Higgs to $WW$ and $ZZ$ gauge boson pairs, as well bottom quarks given by $r_{\mathcal HWW}$, $r_{\mathcal HZZ}$, and $r_{\mathcal Hbb}$, respectively ($r_{\mathcal HXX} \equiv g_{\mathcal HXX}/g^{SM}_{\mathcal HXX}$).~In~\fref{WW}, we show results for $r_{\mathcal HWW}$ (dark green dotted),~$r_{\mathcal HZZ}$ (blue dashed),~and $r_{\mathcal Hbb}$ (black solid) in the $(v_\Delta,\mathcal M)$ plane.~Again we superimpose the region allowed by electroweak precision constraints (pink shaded region).~In the SCTM the Higgs can have couplings to $W$ and $Z$ bosons larger than the ones predicted by the SM (see also~\cite{Logan:2010en,Falkowski:2012vh}), but still well within current experimental bounds~\cite{CMS-PAS-HIG-14-009,ATLAS-CONF-2014-009}.~In particular, at large values of $v_\Delta$, the two couplings can deviate from the SM prediction by as much as $(5-10)\%$ for our chosen parameter point.~Such a deviation could possibly be measured at a high luminosity LHC~\cite{Peskin:2013xra,ATLAS:2013hta,CMS:2013xfa,Brock:2014tja}.~This is in contrast to models with only additional Higgs doublets and singlets, which can only reduce the Higgs couplings to gauge bosons.~This has interesting implications for trying to extract the total width of the 125 GeV Higgs boson without making the theoretical assumption $r_{\mathcal HWW},\,r_{\mathcal HZZ}\leq 1$ (see e.g.~\cite{Dobrescu:2012td,Belanger:2013xza}).~We also see in~\fref{WW} that, for this parameter point, the Higgs coupling to bottom quarks is only mildly modified, with respect to the SM.
\begin{figure}[htb]
\begin{center}
\includegraphics[width=85mm]{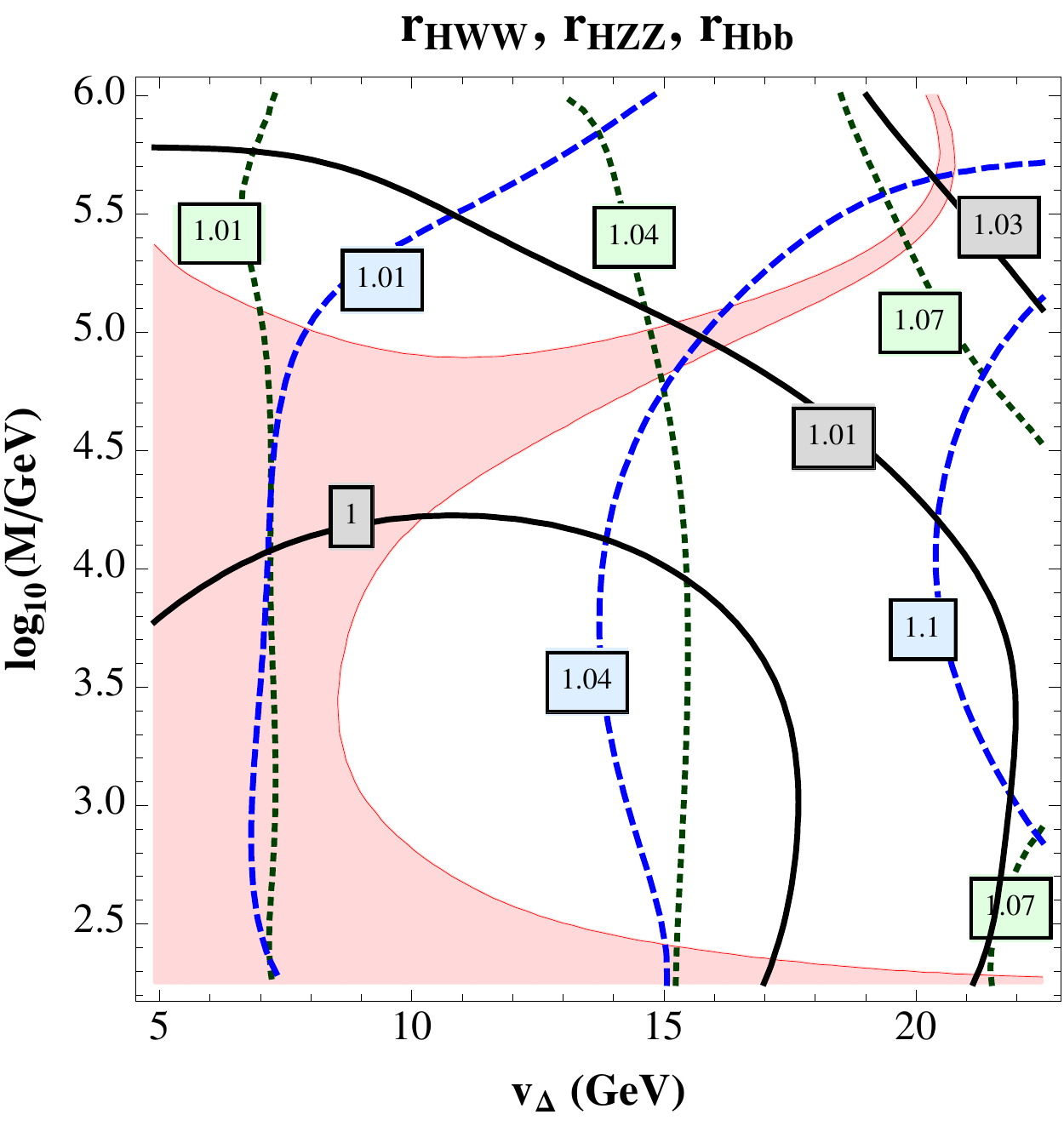}
\end{center}
\caption{\it Contours of $r_{\mathcal HWW}$ (dark green dotted), $r_{\mathcal HZZ}$ (blue dashed), and $r_{\mathcal Hbb}$ (black solid) in the $(v_\Delta,\mathcal M)$ plane for the values of the parameters given in~\eref{valores} and $m_{1/2}=1.2$~TeV.}
\label{fig:WW}
\end{figure}

It is also interesting to examine the ratio of of the normalized couplings $r_{\mathcal HWW}/r_{\mathcal HZZ} \equiv \lambda_{WZ}$~\cite{CMS-PAS-HIG-14-009,ATLAS-CONF-2014-009}, since it is a direct measure of the violation of custodial symmetry induced by the RGE running.~In the SM and in the MSSM, custodial symmetry implies $\lambda_{WZ} = 1$, but in the SCTM it is possible to have deviations from this universal relation.~In~\fref{RWWZZvsMgaugino}, we show the quantity $\lambda_{WZ} -1$ as a function of the gaugino mass $m_{1/2}$ and $v_\Delta$ along the $2v_\phi^2=v_\chi^2+v_\psi^2$ (i.e.~$\tan\theta_0 = 1$ yielding $\Delta\rho=0$) direction for parameter values given in~\eref{valores} and $\mathcal M = 850$~TeV.~Since in the SCTM the ratio $\lambda_{WZ}$ is a function of all three vacuum angles ($\beta, \theta_0, \theta_1$) it will be in general different from one, even in the direction $2v_\phi^2=v_\chi^2+v_\psi^2$, on which $\Delta\rho=0$.~At large values of $v_\Delta$ deviations from universality as large as $\sim (10-15)\%$ are achievable.~This is well within present experimental constraints~\cite{CMS-PAS-HIG-14-009,ATLAS-CONF-2014-009} and potentially observable at a HL-LHC~\cite{Peskin:2013xra,ATLAS:2013hta,CMS:2013xfa,Brock:2014tja}.
\vspace*{-.168cm}
\begin{figure}[htb]
\begin{center}
\includegraphics[width=88mm]{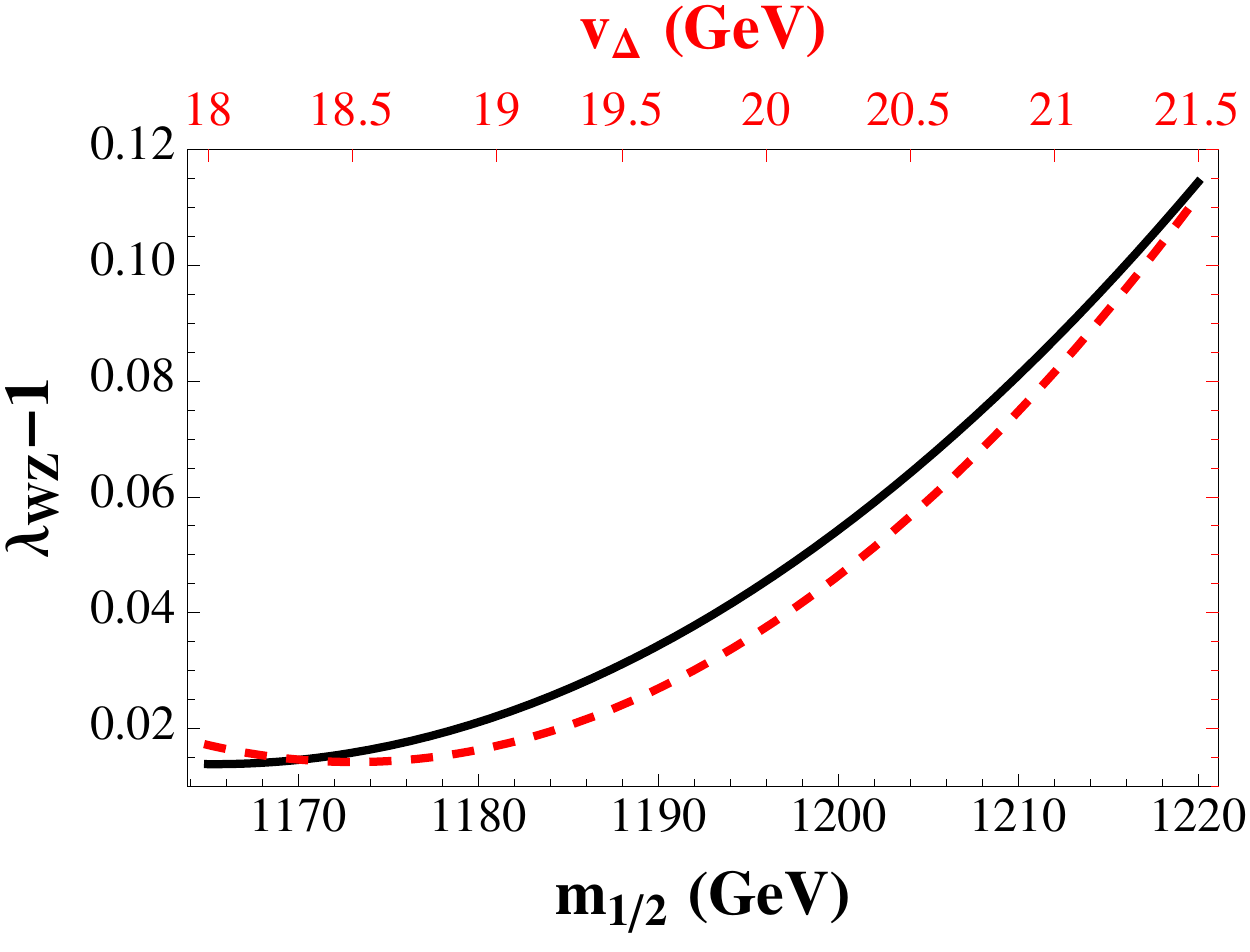}
\end{center}
\caption{\it Deviation from the universal condition $\lambda_{WZ} = 1$ along the $2v_\phi^2=v_\chi^2+v_\psi^2$ direction (or $\tan\theta_0 = 1$, which provides $\Delta\rho=0$) as a function of $m_{1/2}$ (black solid line) and $v_\Delta$ (red dashed line) for parameter values given in~\eref{valores} and $\mathcal M=850$ TeV.}
\label{fig:RWWZZvsMgaugino}
\end{figure}

Of course there are many additional Higgs observables that could be used to test the SCTM.~Generically, the particle spectrum has several TeV-scale charged particles which can contribute to the $\mathcal H\gamma\gamma$ decay width.~These particles will also modify the $\mathcal H \to 4\ell$ and $\mathcal H \to 2\ell\gamma$ decays, which could be used to probe the underlying CP properties of the model~\cite{DeRujula:2010ys,Christensen:2010pf,Anderson:2013afp,Gainer:2013rxa,Chen:2014gka,Chen:2014ona,Falkowski:2014ffa}.

Furthermore, the model will be tested by the direct searches for the additional scalars and fermions arising in the spectrum.~Particularly interesting signatures are the decays of the doubly charged Higgs scalars to $W^\pm W^\pm$~\cite{Godfrey:2010qb,Englert:2013zpa,Kanemura:2014goa} and the decay of the singly charged scalars to $W^{\pm}Z$, a decay found only in models with larger than doublet representations~\cite{Gunion:1989ci}.~Additionally, in the SCTM the doubly charged Higgsino will decay to same sign $W$ boson pairs plus missing energy.~In particular, a doubly charged fermion with a mass near that of the doubly charged scalar would be a strong hint of the SCTM.~A precise determination of the LHC sensitivity to these signals deserves a more careful treatment which is beyond the scope of this work. 

\section{Discussion and Conclusions}\label{sec:disc}
We have constructed a model dubbed the supersymmetric custodial Higgs triplets model (SCTM) with an extended Higgs sector which includes electroweak triplets that can significantly contribute to EWSB while satisfying the relevant experimental constraints coming from electroweak precision data and LHC measurements.~We have discussed how this model can address the naturalness problems associated with the well-known Georgi-Machacek (GM) model.~In particular, this theory is free both from the quadratic divergences found in the GM model and from the need to arbitrarily set the scale at which the global $SU(2)_L\otimes SU(2)_R$ invariance holds at the electroweak scale, in order to obtain $\rho \approx 1$.

By utilizing an extra VEV direction, which itself is a consequence of supersymmetry and anomaly cancellation, we have shown that the scale $\mathcal{M}$ at which $SU(2)_L\otimes SU(2)_R$ invariance holds can be significantly higher than the electroweak scale.~In particular, we find that scales $\mathcal M$ well above $100$ TeV and triplet contributions to EWSB as large as $15\%$ can easily be accommodated.~We have also argued that in the SCTM,~$\mathcal{M}$ is most naturally identified with the messenger scale, at which supersymmetry breaking is transmitted to the observable sector, leading to a connection between the experimentally measured value of $\rho$ and the supersymmetry breaking scale.~With this identification, we have demonstrated that, once the $SU(2)_L\otimes SU(2)_R$ boundary conditions are specified at the scale $\mathcal M$, then for a given triplet VEV, the tree-level value of $\rho$ can be predicted through renormalization group evolution.

At the same time we have demonstrated that the SCTM can easily give large tree-level \emph{and} one-loop contributions to the Higgs mass.~This allows for reproducing the measured Higgs mass even with small trilinear terms and top squarks with mass below $1$~TeV.

Finally, we have discussed a number smoking guns of the SCTM including the possibility of enhanced Higgs coupling to $WW$ and $ZZ$, a feature shared among all Higgs triplet models.~We have also examined the possibility of departure from the universal relation of the Higgs couplings to $W$ and $Z$ bosons ($r_{\mathcal H WW} = r_{\mathcal H ZZ}$), while still obtaining $\rho \approx 1$, that is a unique feature of the model and a measure of custodial symmetry violation at the electroweak scale.

There are still many potential avenues of exploration for the SCTM left open in the present Letter.~For example, it is interesting to consider potential UV completions which provide a mechanism for supersymmetry breaking and generating the $SU(2)_L\otimes SU(2)_R$ invariant Higgs sector.~Furthermore, the one-loop corrections and potential threshold effects in our analysis of EW precision observables, as well as a dedicated LHC study, may provide additional insight to the SCTM.~We leave these avenues of exploration to ongoing work \cite{preparation}.

\bigskip

{\em Acknowledgements:} 
We would like to thank Ian Low and Joe Lykken for useful conversations in the early stages of this work.~We would like to thank Wolfgang Altmannshofer, Patrick Meade, Javi Serra, and Carlos Wagner for useful comments on the draft.~The work of M.G.-P.~and M.Q.~is partly supported by the Spanish Consolider-Ingenio 2010 Programme CPAN under Grants CSD2007-00042, CICYT-FEDER-FPA2011-25948 and by \textit{Secretaria d'Universitats i Recerca del Departament d'Economia i Coneixement de la Generalitat de Catalunya} under Grant 2014 SGR 1450.~The work of R.V.M.~is supported by the ERC Advanced Grant Higgs@LHC.~The work of T.-T.Y.~is supported in part by the National Science Foundation under Grant PHY-PHY-0969739.~R.V.~wishes to thank LPT-Orsay for its hospitality during the completion of this work.~S.G.~and T.-T.Y.~wish to thank the Aspen Center for Physics, which operates under the NSF Grant 1066293, for hospitality during the completion of this work.~S.G.~would like to thank the SLAC theory group and the CFHEP for hospitality and partial support.~Research at Perimeter Institute is supported by the Government of Canada through Industry Canada
and by the Province of Ontario through the Ministry of
Economic Development \& Innovation.


\bibliographystyle{apsrev}
\bibliography{short}

\begin{thebibliography}{63}
\expandafter\ifx\csname natexlab\endcsname\relax\def\natexlab#1{#1}\fi
\expandafter\ifx\csname bibnamefont\endcsname\relax
  \def\bibnamefont#1{#1}\fi
\expandafter\ifx\csname bibfnamefont\endcsname\relax
  \def\bibfnamefont#1{#1}\fi
\expandafter\ifx\csname citenamefont\endcsname\relax
  \def\citenamefont#1{#1}\fi
\expandafter\ifx\csname url\endcsname\relax
  \def\url#1{\texttt{#1}}\fi
\expandafter\ifx\csname urlprefix\endcsname\relax\def\urlprefix{URL }\fi
\providecommand{\bibinfo}[2]{#2}
\providecommand{\eprint}[2][]{\url{#2}}

\bibitem[{\citenamefont{Chatrchyan et~al.}(2012)}]{Chatrchyan:2012ufa}
\bibinfo{author}{\bibfnamefont{S.}~\bibnamefont{Chatrchyan}}
  \bibnamefont{et~al.} (\bibinfo{collaboration}{CMS Collaboration}),
  \bibinfo{journal}{Phys.Lett.} \textbf{\bibinfo{volume}{B716}},
  \bibinfo{pages}{30} (\bibinfo{year}{2012}), \eprint{1207.7235}.

\bibitem[{\citenamefont{Aad et~al.}(2012)}]{Aad:2012tfa}
\bibinfo{author}{\bibfnamefont{G.}~\bibnamefont{Aad}} \bibnamefont{et~al.}
  (\bibinfo{collaboration}{ATLAS Collaboration}), \bibinfo{journal}{Phys.Lett.}
  \textbf{\bibinfo{volume}{B716}}, \bibinfo{pages}{1} (\bibinfo{year}{2012}),
  \eprint{1207.7214}.

\bibitem[{\citenamefont{Falkowski et~al.}(2013)\citenamefont{Falkowski, Riva,
  and Urbano}}]{Falkowski:2013dza}
\bibinfo{author}{\bibfnamefont{A.}~\bibnamefont{Falkowski}},
  \bibinfo{author}{\bibfnamefont{F.}~\bibnamefont{Riva}}, \bibnamefont{and}
  \bibinfo{author}{\bibfnamefont{A.}~\bibnamefont{Urbano}}
  (\bibinfo{year}{2013}), \eprint{1303.1812}.

\bibitem[{CMS(2014)}]{CMS-PAS-HIG-14-009}
\bibinfo{type}{Tech. Rep.} \bibinfo{number}{CMS-PAS-HIG-14-009},
  \bibinfo{institution}{CMS Collaboration}, \bibinfo{address}{Geneva}
  (\bibinfo{year}{2014}).

\bibitem[{ATL(2014)}]{ATLAS-CONF-2014-009}
\bibinfo{type}{Tech. Rep.} \bibinfo{number}{ATLAS-CONF-2014-009},
  \bibinfo{institution}{ATLAS Collaboration}, \bibinfo{address}{Geneva}
  (\bibinfo{year}{2014}).

\bibitem[{\citenamefont{Low and Lykken}(2010)}]{Low:2010jp}
\bibinfo{author}{\bibfnamefont{I.}~\bibnamefont{Low}} \bibnamefont{and}
  \bibinfo{author}{\bibfnamefont{J.}~\bibnamefont{Lykken}},
  \bibinfo{journal}{JHEP} \textbf{\bibinfo{volume}{1010}}, \bibinfo{pages}{053}
  (\bibinfo{year}{2010}), \eprint{1005.0872}.

\bibitem[{\citenamefont{Beringer et~al.}(2012)}]{Beringer:1900zz}
\bibinfo{author}{\bibfnamefont{J.}~\bibnamefont{Beringer}} \bibnamefont{et~al.}
  (\bibinfo{collaboration}{Particle Data Group}), \bibinfo{journal}{Phys.Rev.}
  \textbf{\bibinfo{volume}{D86}}, \bibinfo{pages}{010001}
  (\bibinfo{year}{2012}).

\bibitem[{\citenamefont{Georgi and Machacek}(1985)}]{Georgi:1985nv}
\bibinfo{author}{\bibfnamefont{H.}~\bibnamefont{Georgi}} \bibnamefont{and}
  \bibinfo{author}{\bibfnamefont{M.}~\bibnamefont{Machacek}},
  \bibinfo{journal}{Nucl.Phys.} \textbf{\bibinfo{volume}{B262}},
  \bibinfo{pages}{463} (\bibinfo{year}{1985}).

\bibitem[{\citenamefont{Chanowitz and Golden}(1985)}]{Chanowitz:1985ug}
\bibinfo{author}{\bibfnamefont{M.~S.} \bibnamefont{Chanowitz}}
  \bibnamefont{and} \bibinfo{author}{\bibfnamefont{M.}~\bibnamefont{Golden}},
  \bibinfo{journal}{Phys.Lett.} \textbf{\bibinfo{volume}{B165}},
  \bibinfo{pages}{105} (\bibinfo{year}{1985}).

\bibitem[{\citenamefont{Gunion et~al.}(1990)\citenamefont{Gunion, Vega, and
  Wudka}}]{Gunion:1989ci}
\bibinfo{author}{\bibfnamefont{J.}~\bibnamefont{Gunion}},
  \bibinfo{author}{\bibfnamefont{R.}~\bibnamefont{Vega}}, \bibnamefont{and}
  \bibinfo{author}{\bibfnamefont{J.}~\bibnamefont{Wudka}},
  \bibinfo{journal}{Phys.Rev.} \textbf{\bibinfo{volume}{D42}},
  \bibinfo{pages}{1673} (\bibinfo{year}{1990}).

\bibitem[{\citenamefont{Gunion et~al.}(1991)\citenamefont{Gunion, Vega, and
  Wudka}}]{Gunion:1990dt}
\bibinfo{author}{\bibfnamefont{J.}~\bibnamefont{Gunion}},
  \bibinfo{author}{\bibfnamefont{R.}~\bibnamefont{Vega}}, \bibnamefont{and}
  \bibinfo{author}{\bibfnamefont{J.}~\bibnamefont{Wudka}},
  \bibinfo{journal}{Phys.Rev.} \textbf{\bibinfo{volume}{D43}},
  \bibinfo{pages}{2322} (\bibinfo{year}{1991}).

\bibitem[{\citenamefont{Godfrey and Moats}(2010)}]{Godfrey:2010qb}
\bibinfo{author}{\bibfnamefont{S.}~\bibnamefont{Godfrey}} \bibnamefont{and}
  \bibinfo{author}{\bibfnamefont{K.}~\bibnamefont{Moats}},
  \bibinfo{journal}{Phys.Rev.} \textbf{\bibinfo{volume}{D81}},
  \bibinfo{pages}{075026} (\bibinfo{year}{2010}), \eprint{1003.3033}.

\bibitem[{\citenamefont{Englert
  et~al.}(2013{\natexlab{a}})\citenamefont{Englert, Re, and
  Spannowsky}}]{Englert:2013zpa}
\bibinfo{author}{\bibfnamefont{C.}~\bibnamefont{Englert}},
  \bibinfo{author}{\bibfnamefont{E.}~\bibnamefont{Re}}, \bibnamefont{and}
  \bibinfo{author}{\bibfnamefont{M.}~\bibnamefont{Spannowsky}},
  \bibinfo{journal}{Phys.Rev.} \textbf{\bibinfo{volume}{D87}},
  \bibinfo{pages}{095014} (\bibinfo{year}{2013}{\natexlab{a}}),
  \eprint{1302.6505}.

\bibitem[{\citenamefont{Aoki and Kanemura}(2008)}]{Aoki:2007ah}
\bibinfo{author}{\bibfnamefont{M.}~\bibnamefont{Aoki}} \bibnamefont{and}
  \bibinfo{author}{\bibfnamefont{S.}~\bibnamefont{Kanemura}},
  \bibinfo{journal}{Phys.Rev.} \textbf{\bibinfo{volume}{D77}},
  \bibinfo{pages}{095009} (\bibinfo{year}{2008}), \eprint{0712.4053}.

\bibitem[{\citenamefont{Carmi et~al.}(2012)\citenamefont{Carmi, Falkowski,
  Kuflik, Volansky, and Zupan}}]{Carmi:2012in}
\bibinfo{author}{\bibfnamefont{D.}~\bibnamefont{Carmi}},
  \bibinfo{author}{\bibfnamefont{A.}~\bibnamefont{Falkowski}},
  \bibinfo{author}{\bibfnamefont{E.}~\bibnamefont{Kuflik}},
  \bibinfo{author}{\bibfnamefont{T.}~\bibnamefont{Volansky}}, \bibnamefont{and}
  \bibinfo{author}{\bibfnamefont{J.}~\bibnamefont{Zupan}},
  \bibinfo{journal}{JHEP} \textbf{\bibinfo{volume}{1210}}, \bibinfo{pages}{196}
  (\bibinfo{year}{2012}), \eprint{1207.1718}.

\bibitem[{\citenamefont{Chiang and Yagyu}(2013)}]{Chiang:2012cn}
\bibinfo{author}{\bibfnamefont{C.-W.} \bibnamefont{Chiang}} \bibnamefont{and}
  \bibinfo{author}{\bibfnamefont{K.}~\bibnamefont{Yagyu}},
  \bibinfo{journal}{JHEP} \textbf{\bibinfo{volume}{1301}}, \bibinfo{pages}{026}
  (\bibinfo{year}{2013}), \eprint{1211.2658}.

\bibitem[{\citenamefont{Kanemura et~al.}(2013)\citenamefont{Kanemura, Kikuchi,
  and Yagyu}}]{Kanemura:2013mc}
\bibinfo{author}{\bibfnamefont{S.}~\bibnamefont{Kanemura}},
  \bibinfo{author}{\bibfnamefont{M.}~\bibnamefont{Kikuchi}}, \bibnamefont{and}
  \bibinfo{author}{\bibfnamefont{K.}~\bibnamefont{Yagyu}},
  \bibinfo{journal}{Phys.Rev.} \textbf{\bibinfo{volume}{D88}},
  \bibinfo{pages}{015020} (\bibinfo{year}{2013}), \eprint{1301.7303}.

\bibitem[{\citenamefont{Chang et~al.}(2012)\citenamefont{Chang, Newby, Raj, and
  Wanotayaroj}}]{Chang:2012gn}
\bibinfo{author}{\bibfnamefont{S.}~\bibnamefont{Chang}},
  \bibinfo{author}{\bibfnamefont{C.~A.} \bibnamefont{Newby}},
  \bibinfo{author}{\bibfnamefont{N.}~\bibnamefont{Raj}}, \bibnamefont{and}
  \bibinfo{author}{\bibfnamefont{C.}~\bibnamefont{Wanotayaroj}},
  \bibinfo{journal}{Phys.Rev.} \textbf{\bibinfo{volume}{D86}},
  \bibinfo{pages}{095015} (\bibinfo{year}{2012}), \eprint{1207.0493}.

\bibitem[{\citenamefont{Chiang}(2013)}]{Chiang:2013bqa}
\bibinfo{author}{\bibfnamefont{C.-W.} \bibnamefont{Chiang}}
  (\bibinfo{year}{2013}), \eprint{1305.1006}.

\bibitem[{\citenamefont{Killick et~al.}(2013)\citenamefont{Killick, Kumar, and
  Logan}}]{Killick:2013mya}
\bibinfo{author}{\bibfnamefont{R.}~\bibnamefont{Killick}},
  \bibinfo{author}{\bibfnamefont{K.}~\bibnamefont{Kumar}}, \bibnamefont{and}
  \bibinfo{author}{\bibfnamefont{H.~E.} \bibnamefont{Logan}},
  \bibinfo{journal}{Phys.Rev.} \textbf{\bibinfo{volume}{D88}},
  \bibinfo{pages}{033015} (\bibinfo{year}{2013}), \eprint{1305.7236}.

\bibitem[{\citenamefont{Belanger et~al.}(2013)\citenamefont{Belanger, Dumont,
  Ellwanger, Gunion, and Kraml}}]{Belanger:2013xza}
\bibinfo{author}{\bibfnamefont{G.}~\bibnamefont{Belanger}},
  \bibinfo{author}{\bibfnamefont{B.}~\bibnamefont{Dumont}},
  \bibinfo{author}{\bibfnamefont{U.}~\bibnamefont{Ellwanger}},
  \bibinfo{author}{\bibfnamefont{J.}~\bibnamefont{Gunion}}, \bibnamefont{and}
  \bibinfo{author}{\bibfnamefont{S.}~\bibnamefont{Kraml}},
  \bibinfo{journal}{Phys.Rev.} \textbf{\bibinfo{volume}{D88}},
  \bibinfo{pages}{075008} (\bibinfo{year}{2013}), \eprint{1306.2941}.

\bibitem[{\citenamefont{Englert
  et~al.}(2013{\natexlab{b}})\citenamefont{Englert, Re, and
  Spannowsky}}]{Englert:2013wga}
\bibinfo{author}{\bibfnamefont{C.}~\bibnamefont{Englert}},
  \bibinfo{author}{\bibfnamefont{E.}~\bibnamefont{Re}}, \bibnamefont{and}
  \bibinfo{author}{\bibfnamefont{M.}~\bibnamefont{Spannowsky}},
  \bibinfo{journal}{Phys.Rev.} \textbf{\bibinfo{volume}{D88}},
  \bibinfo{pages}{035024} (\bibinfo{year}{2013}{\natexlab{b}}),
  \eprint{1306.6228}.

\bibitem[{\citenamefont{Chiang et~al.}(2014)\citenamefont{Chiang, Kanemura, and
  Yagyu}}]{Chiang:2014bia}
\bibinfo{author}{\bibfnamefont{C.-W.} \bibnamefont{Chiang}},
  \bibinfo{author}{\bibfnamefont{S.}~\bibnamefont{Kanemura}}, \bibnamefont{and}
  \bibinfo{author}{\bibfnamefont{K.}~\bibnamefont{Yagyu}}
  (\bibinfo{year}{2014}), \eprint{1407.5053}.

\bibitem[{\citenamefont{Tsao}(1980)}]{Tsao:1980em}
\bibinfo{author}{\bibfnamefont{H.-S.} \bibnamefont{Tsao}}, p.
  \bibinfo{pages}{1240} (\bibinfo{year}{1980}), \eprint{Proc. Guangzhou Conf.}

\bibitem[{\citenamefont{Gunion et~al.}(2000)\citenamefont{Gunion, Haber, Kane,
  and Dawson}}]{Gunion:1989we}
\bibinfo{author}{\bibfnamefont{J.~F.} \bibnamefont{Gunion}},
  \bibinfo{author}{\bibfnamefont{H.~E.} \bibnamefont{Haber}},
  \bibinfo{author}{\bibfnamefont{G.~L.} \bibnamefont{Kane}}, \bibnamefont{and}
  \bibinfo{author}{\bibfnamefont{S.}~\bibnamefont{Dawson}},
  \bibinfo{journal}{Front.Phys.} \textbf{\bibinfo{volume}{80}},
  \bibinfo{pages}{1} (\bibinfo{year}{2000}).

\bibitem[{\citenamefont{Hisano and Tsumura}(2013)}]{Hisano:2013sn}
\bibinfo{author}{\bibfnamefont{J.}~\bibnamefont{Hisano}} \bibnamefont{and}
  \bibinfo{author}{\bibfnamefont{K.}~\bibnamefont{Tsumura}},
  \bibinfo{journal}{Phys.Rev.} \textbf{\bibinfo{volume}{D87}},
  \bibinfo{pages}{053004} (\bibinfo{year}{2013}), \eprint{1301.6455}.

\bibitem[{\citenamefont{Alvarado et~al.}(2014)\citenamefont{Alvarado, Lehman,
  and Ostdiek}}]{Alvarado:2014jva}
\bibinfo{author}{\bibfnamefont{C.}~\bibnamefont{Alvarado}},
  \bibinfo{author}{\bibfnamefont{L.}~\bibnamefont{Lehman}}, \bibnamefont{and}
  \bibinfo{author}{\bibfnamefont{B.}~\bibnamefont{Ostdiek}},
  \bibinfo{journal}{JHEP} \textbf{\bibinfo{volume}{2014}}, \bibinfo{pages}{150}
  (\bibinfo{year}{2014}), \eprint{1404.3208}.

\bibitem[{\citenamefont{Cort et~al.}(2013)\citenamefont{Cort, Garcia, and
  Quiros}}]{Cort:2013foa}
\bibinfo{author}{\bibfnamefont{L.}~\bibnamefont{Cort}},
  \bibinfo{author}{\bibfnamefont{M.}~\bibnamefont{Garcia}}, \bibnamefont{and}
  \bibinfo{author}{\bibfnamefont{M.}~\bibnamefont{Quiros}},
  \bibinfo{journal}{Phys.Rev.} \textbf{\bibinfo{volume}{D88}},
  \bibinfo{pages}{075010} (\bibinfo{year}{2013}), \eprint{1308.4025}.

\bibitem[{\citenamefont{Garcia et~al.}(In preparation)\citenamefont{Garcia,
  Gori, Quiros, Vega, Vega-Morales, and Yu}}]{preparation}
\bibinfo{author}{\bibfnamefont{M.}~\bibnamefont{Garcia}},
  \bibinfo{author}{\bibfnamefont{S.}~\bibnamefont{Gori}},
  \bibinfo{author}{\bibfnamefont{M.}~\bibnamefont{Quiros}},
  \bibinfo{author}{\bibfnamefont{R.}~\bibnamefont{Vega}},
  \bibinfo{author}{\bibfnamefont{R.}~\bibnamefont{Vega-Morales}},
  \bibnamefont{and} \bibinfo{author}{\bibfnamefont{T.-T.} \bibnamefont{Yu}}
  (\bibinfo{year}{In preparation}).

\bibitem[{\citenamefont{Drees and Hagiwara}(1990)}]{Drees:1990dx}
\bibinfo{author}{\bibfnamefont{M.}~\bibnamefont{Drees}} \bibnamefont{and}
  \bibinfo{author}{\bibfnamefont{K.}~\bibnamefont{Hagiwara}},
  \bibinfo{journal}{Phys.Rev.} \textbf{\bibinfo{volume}{D42}},
  \bibinfo{pages}{1709} (\bibinfo{year}{1990}).

\bibitem[{\citenamefont{Fileviez~Perez and
  Spinner}(2013{\natexlab{a}})}]{FileviezPerez:2012ab}
\bibinfo{author}{\bibfnamefont{P.}~\bibnamefont{Fileviez~Perez}}
  \bibnamefont{and} \bibinfo{author}{\bibfnamefont{S.}~\bibnamefont{Spinner}},
  \bibinfo{journal}{Phys.Rev.} \textbf{\bibinfo{volume}{D87}},
  \bibinfo{pages}{031702} (\bibinfo{year}{2013}{\natexlab{a}}),
  \eprint{1211.1025}.

\bibitem[{\citenamefont{Fileviez~Perez and
  Spinner}(2013{\natexlab{b}})}]{FileviezPerez:2012gg}
\bibinfo{author}{\bibfnamefont{P.}~\bibnamefont{Fileviez~Perez}}
  \bibnamefont{and} \bibinfo{author}{\bibfnamefont{S.}~\bibnamefont{Spinner}},
  \bibinfo{journal}{Phys.Lett.} \textbf{\bibinfo{volume}{B723}},
  \bibinfo{pages}{371} (\bibinfo{year}{2013}{\natexlab{b}}),
  \eprint{1209.5769}.

\bibitem[{\citenamefont{Delgado et~al.}(2012)\citenamefont{Delgado, Nardini,
  and Quiros}}]{Delgado:2012sm}
\bibinfo{author}{\bibfnamefont{A.}~\bibnamefont{Delgado}},
  \bibinfo{author}{\bibfnamefont{G.}~\bibnamefont{Nardini}}, \bibnamefont{and}
  \bibinfo{author}{\bibfnamefont{M.}~\bibnamefont{Quiros}},
  \bibinfo{journal}{Phys.Rev.} \textbf{\bibinfo{volume}{D86}},
  \bibinfo{pages}{115010} (\bibinfo{year}{2012}), \eprint{1207.6596}.

\bibitem[{\citenamefont{Delgado et~al.}(2013)\citenamefont{Delgado, Nardini,
  and Quiros}}]{Delgado:2013zfa}
\bibinfo{author}{\bibfnamefont{A.}~\bibnamefont{Delgado}},
  \bibinfo{author}{\bibfnamefont{G.}~\bibnamefont{Nardini}}, \bibnamefont{and}
  \bibinfo{author}{\bibfnamefont{M.}~\bibnamefont{Quiros}},
  \bibinfo{journal}{JHEP} \textbf{\bibinfo{volume}{1307}}, \bibinfo{pages}{054}
  (\bibinfo{year}{2013}), \eprint{1303.0800}.

\bibitem[{\citenamefont{Hartling et~al.}(2014)\citenamefont{Hartling, Kumar,
  and Logan}}]{Hartling:2014zca}
\bibinfo{author}{\bibfnamefont{K.}~\bibnamefont{Hartling}},
  \bibinfo{author}{\bibfnamefont{K.}~\bibnamefont{Kumar}}, \bibnamefont{and}
  \bibinfo{author}{\bibfnamefont{H.~E.} \bibnamefont{Logan}},
  \bibinfo{journal}{Phys.Rev.} \textbf{\bibinfo{volume}{D90}},
  \bibinfo{pages}{015007} (\bibinfo{year}{2014}), \eprint{1404.2640}.

\bibitem[{\citenamefont{Espinosa and
  Quiros}(1992{\natexlab{a}})}]{Espinosa:1991gr}
\bibinfo{author}{\bibfnamefont{J.}~\bibnamefont{Espinosa}} \bibnamefont{and}
  \bibinfo{author}{\bibfnamefont{M.}~\bibnamefont{Quiros}},
  \bibinfo{journal}{Phys.Lett.} \textbf{\bibinfo{volume}{B279}},
  \bibinfo{pages}{92} (\bibinfo{year}{1992}{\natexlab{a}}).

\bibitem[{\citenamefont{Espinosa and
  Quiros}(1992{\natexlab{b}})}]{Espinosa:1991wt}
\bibinfo{author}{\bibfnamefont{J.}~\bibnamefont{Espinosa}} \bibnamefont{and}
  \bibinfo{author}{\bibfnamefont{M.}~\bibnamefont{Quiros}},
  \bibinfo{journal}{Nucl.Phys.} \textbf{\bibinfo{volume}{B384}},
  \bibinfo{pages}{113} (\bibinfo{year}{1992}{\natexlab{b}}).

\bibitem[{\citenamefont{Espinosa and Quiros}(1993)}]{Espinosa:1992hp}
\bibinfo{author}{\bibfnamefont{J.}~\bibnamefont{Espinosa}} \bibnamefont{and}
  \bibinfo{author}{\bibfnamefont{M.}~\bibnamefont{Quiros}},
  \bibinfo{journal}{Phys.Lett.} \textbf{\bibinfo{volume}{B302}},
  \bibinfo{pages}{51} (\bibinfo{year}{1993}), \eprint{hep-ph/9212305}.

\bibitem[{\citenamefont{Kane et~al.}(1993)\citenamefont{Kane, Kolda, and
  Wells}}]{Kane:1992kq}
\bibinfo{author}{\bibfnamefont{G.~L.} \bibnamefont{Kane}},
  \bibinfo{author}{\bibfnamefont{C.~F.} \bibnamefont{Kolda}}, \bibnamefont{and}
  \bibinfo{author}{\bibfnamefont{J.~D.} \bibnamefont{Wells}},
  \bibinfo{journal}{Phys.Rev.Lett.} \textbf{\bibinfo{volume}{70}},
  \bibinfo{pages}{2686} (\bibinfo{year}{1993}), \eprint{hep-ph/9210242}.

\bibitem[{\citenamefont{Quiros and Espinosa}(1998)}]{Quiros:1998bz}
\bibinfo{author}{\bibfnamefont{M.}~\bibnamefont{Quiros}} \bibnamefont{and}
  \bibinfo{author}{\bibfnamefont{J.~R.} \bibnamefont{Espinosa}},
  \bibinfo{journal}{6th International Symposium on Particles, strings and
  cosmology (PASCOS 1998)} pp. \bibinfo{pages}{513--520}
  (\bibinfo{year}{1998}), \eprint{hep-ph/9809269}.

\bibitem[{\citenamefont{Basak and Mohanty}(2012)}]{Basak:2012bd}
\bibinfo{author}{\bibfnamefont{T.}~\bibnamefont{Basak}} \bibnamefont{and}
  \bibinfo{author}{\bibfnamefont{S.}~\bibnamefont{Mohanty}},
  \bibinfo{journal}{Phys.Rev.} \textbf{\bibinfo{volume}{D86}},
  \bibinfo{pages}{075031} (\bibinfo{year}{2012}), \eprint{1204.6592}.

\bibitem[{\citenamefont{Bandyopadhyay et~al.}(2013)\citenamefont{Bandyopadhyay,
  Huitu, and Sabanci}}]{Bandyopadhyay:2013lca}
\bibinfo{author}{\bibfnamefont{P.}~\bibnamefont{Bandyopadhyay}},
  \bibinfo{author}{\bibfnamefont{K.}~\bibnamefont{Huitu}}, \bibnamefont{and}
  \bibinfo{author}{\bibfnamefont{A.}~\bibnamefont{Sabanci}},
  \bibinfo{journal}{JHEP} \textbf{\bibinfo{volume}{1310}}, \bibinfo{pages}{091}
  (\bibinfo{year}{2013}), \eprint{1306.4530}.

\bibitem[{\citenamefont{Bandyopadhyay et~al.}(2014)\citenamefont{Bandyopadhyay,
  Di~Chiara, Huitu, and Keeli}}]{Bandyopadhyay:2014tha}
\bibinfo{author}{\bibfnamefont{P.}~\bibnamefont{Bandyopadhyay}},
  \bibinfo{author}{\bibfnamefont{S.}~\bibnamefont{Di~Chiara}},
  \bibinfo{author}{\bibfnamefont{K.}~\bibnamefont{Huitu}}, \bibnamefont{and}
  \bibinfo{author}{\bibfnamefont{A.~S.} \bibnamefont{Keeli}}
  (\bibinfo{year}{2014}), \eprint{1407.4836}.

\bibitem[{\citenamefont{Draper et~al.}(2012)\citenamefont{Draper, Meade, Reece,
  and Shih}}]{Draper:2011aa}
\bibinfo{author}{\bibfnamefont{P.}~\bibnamefont{Draper}},
  \bibinfo{author}{\bibfnamefont{P.}~\bibnamefont{Meade}},
  \bibinfo{author}{\bibfnamefont{M.}~\bibnamefont{Reece}}, \bibnamefont{and}
  \bibinfo{author}{\bibfnamefont{D.}~\bibnamefont{Shih}},
  \bibinfo{journal}{Phys.Rev.} \textbf{\bibinfo{volume}{D85}},
  \bibinfo{pages}{095007} (\bibinfo{year}{2012}), \eprint{1112.3068}.

\bibitem[{\citenamefont{Feng et~al.}(2013)\citenamefont{Feng, Kant, Profumo,
  and Sanford}}]{Feng:2013tvd}
\bibinfo{author}{\bibfnamefont{J.~L.} \bibnamefont{Feng}},
  \bibinfo{author}{\bibfnamefont{P.}~\bibnamefont{Kant}},
  \bibinfo{author}{\bibfnamefont{S.}~\bibnamefont{Profumo}}, \bibnamefont{and}
  \bibinfo{author}{\bibfnamefont{D.}~\bibnamefont{Sanford}},
  \bibinfo{journal}{Phys.Rev.Lett.} \textbf{\bibinfo{volume}{111}},
  \bibinfo{pages}{131802} (\bibinfo{year}{2013}), \eprint{1306.2318}.

\bibitem[{\citenamefont{Delgado et~al.}(2014)\citenamefont{Delgado, Garcia, and
  Quiros}}]{Delgado:2013gza}
\bibinfo{author}{\bibfnamefont{A.}~\bibnamefont{Delgado}},
  \bibinfo{author}{\bibfnamefont{M.}~\bibnamefont{Garcia}}, \bibnamefont{and}
  \bibinfo{author}{\bibfnamefont{M.}~\bibnamefont{Quiros}},
  \bibinfo{journal}{Phys.Rev.} \textbf{\bibinfo{volume}{D90}},
  \bibinfo{pages}{015016} (\bibinfo{year}{2014}), \eprint{1312.3235}.

\bibitem[{\citenamefont{Draper et~al.}(2014)\citenamefont{Draper, Lee, and
  Wagner}}]{Draper:2013oza}
\bibinfo{author}{\bibfnamefont{P.}~\bibnamefont{Draper}},
  \bibinfo{author}{\bibfnamefont{G.}~\bibnamefont{Lee}}, \bibnamefont{and}
  \bibinfo{author}{\bibfnamefont{C.~E.~M.} \bibnamefont{Wagner}},
  \bibinfo{journal}{Phys.Rev.} \textbf{\bibinfo{volume}{D89}},
  \bibinfo{pages}{055023} (\bibinfo{year}{2014}), \eprint{1312.5743}.

\bibitem[{\citenamefont{Agashe et~al.}(2011)\citenamefont{Agashe, Azatov, Katz,
  and Kim}}]{Agashe:2011ia}
\bibinfo{author}{\bibfnamefont{K.}~\bibnamefont{Agashe}},
  \bibinfo{author}{\bibfnamefont{A.}~\bibnamefont{Azatov}},
  \bibinfo{author}{\bibfnamefont{A.}~\bibnamefont{Katz}}, \bibnamefont{and}
  \bibinfo{author}{\bibfnamefont{D.}~\bibnamefont{Kim}},
  \bibinfo{journal}{Phys.Rev.} \textbf{\bibinfo{volume}{D84}},
  \bibinfo{pages}{115024} (\bibinfo{year}{2011}), \eprint{1109.2842}.

\bibitem[{\citenamefont{Logan and Roy}(2010)}]{Logan:2010en}
\bibinfo{author}{\bibfnamefont{H.~E.} \bibnamefont{Logan}} \bibnamefont{and}
  \bibinfo{author}{\bibfnamefont{M.-A.} \bibnamefont{Roy}},
  \bibinfo{journal}{Phys.Rev.} \textbf{\bibinfo{volume}{D82}},
  \bibinfo{pages}{115011} (\bibinfo{year}{2010}), \eprint{1008.4869}.

\bibitem[{\citenamefont{Falkowski et~al.}(2012)\citenamefont{Falkowski,
  Rychkov, and Urbano}}]{Falkowski:2012vh}
\bibinfo{author}{\bibfnamefont{A.}~\bibnamefont{Falkowski}},
  \bibinfo{author}{\bibfnamefont{S.}~\bibnamefont{Rychkov}}, \bibnamefont{and}
  \bibinfo{author}{\bibfnamefont{A.}~\bibnamefont{Urbano}},
  \bibinfo{journal}{JHEP} \textbf{\bibinfo{volume}{1204}}, \bibinfo{pages}{073}
  (\bibinfo{year}{2012}), \eprint{1202.1532}.

\bibitem[{\citenamefont{Peskin}(2013)}]{Peskin:2013xra}
\bibinfo{author}{\bibfnamefont{M.~E.} \bibnamefont{Peskin}}
  (\bibinfo{year}{2013}), \eprint{1312.4974}.

\bibitem[{{ATLAS Collaboration}()}]{ATLAS:2013hta}
{ATLAS Collaboration} (\bibinfo{year}{2013}), \eprint{1307.7292}.

\bibitem[{CMS Collaboration()}]{CMS:2013xfa}
CMS Collaboration (\bibinfo{year}{2013}), \eprint{1307.7135}.

\bibitem[{\citenamefont{Brock et~al.}(2014)\citenamefont{Brock, Peskin, Agashe,
  Artuso, Campbell et~al.}}]{Brock:2014tja}
\bibinfo{author}{\bibfnamefont{R.}~\bibnamefont{Brock}},
  \bibinfo{author}{\bibfnamefont{M.}~\bibnamefont{Peskin}},
  \bibinfo{author}{\bibfnamefont{K.}~\bibnamefont{Agashe}},
  \bibinfo{author}{\bibfnamefont{M.}~\bibnamefont{Artuso}},
  \bibinfo{author}{\bibfnamefont{J.}~\bibnamefont{Campbell}},
  \bibnamefont{et~al.} (\bibinfo{year}{2014}), \eprint{1401.6081}.

\bibitem[{\citenamefont{Dobrescu and Lykken}(2013)}]{Dobrescu:2012td}
\bibinfo{author}{\bibfnamefont{B.~A.} \bibnamefont{Dobrescu}} \bibnamefont{and}
  \bibinfo{author}{\bibfnamefont{J.~D.} \bibnamefont{Lykken}},
  \bibinfo{journal}{JHEP} \textbf{\bibinfo{volume}{1302}}, \bibinfo{pages}{073}
  (\bibinfo{year}{2013}), \eprint{1210.3342}.

\bibitem[{\citenamefont{De~Rujula et~al.}(2010)\citenamefont{De~Rujula, Lykken,
  Pierini, Rogan, and Spiropulu}}]{DeRujula:2010ys}
\bibinfo{author}{\bibfnamefont{A.}~\bibnamefont{De~Rujula}},
  \bibinfo{author}{\bibfnamefont{J.}~\bibnamefont{Lykken}},
  \bibinfo{author}{\bibfnamefont{M.}~\bibnamefont{Pierini}},
  \bibinfo{author}{\bibfnamefont{C.}~\bibnamefont{Rogan}}, \bibnamefont{and}
  \bibinfo{author}{\bibfnamefont{M.}~\bibnamefont{Spiropulu}},
  \bibinfo{journal}{Phys.Rev.} \textbf{\bibinfo{volume}{D82}},
  \bibinfo{pages}{013003} (\bibinfo{year}{2010}), \eprint{1001.5300}.

\bibitem[{\citenamefont{Christensen et~al.}(2010)\citenamefont{Christensen,
  Han, and Li}}]{Christensen:2010pf}
\bibinfo{author}{\bibfnamefont{N.~D.} \bibnamefont{Christensen}},
  \bibinfo{author}{\bibfnamefont{T.}~\bibnamefont{Han}}, \bibnamefont{and}
  \bibinfo{author}{\bibfnamefont{Y.}~\bibnamefont{Li}},
  \bibinfo{journal}{Phys.Lett.} \textbf{\bibinfo{volume}{B693}},
  \bibinfo{pages}{28} (\bibinfo{year}{2010}), \eprint{1005.5393}.

\bibitem[{\citenamefont{Anderson et~al.}(2014)\citenamefont{Anderson,
  Bolognesi, Caola, Gao, Gritsan et~al.}}]{Anderson:2013afp}
\bibinfo{author}{\bibfnamefont{I.}~\bibnamefont{Anderson}},
  \bibinfo{author}{\bibfnamefont{S.}~\bibnamefont{Bolognesi}},
  \bibinfo{author}{\bibfnamefont{F.}~\bibnamefont{Caola}},
  \bibinfo{author}{\bibfnamefont{Y.}~\bibnamefont{Gao}},
  \bibinfo{author}{\bibfnamefont{A.~V.} \bibnamefont{Gritsan}},
  \bibnamefont{et~al.}, \bibinfo{journal}{Phys.Rev.}
  \textbf{\bibinfo{volume}{D89}}, \bibinfo{pages}{035007}
  (\bibinfo{year}{2014}), \eprint{1309.4819}.

\bibitem[{\citenamefont{Gainer et~al.}(2013)\citenamefont{Gainer, Lykken,
  Matchev, Mrenna, and Park}}]{Gainer:2013rxa}
\bibinfo{author}{\bibfnamefont{J.~S.} \bibnamefont{Gainer}},
  \bibinfo{author}{\bibfnamefont{J.}~\bibnamefont{Lykken}},
  \bibinfo{author}{\bibfnamefont{K.~T.} \bibnamefont{Matchev}},
  \bibinfo{author}{\bibfnamefont{S.}~\bibnamefont{Mrenna}}, \bibnamefont{and}
  \bibinfo{author}{\bibfnamefont{M.}~\bibnamefont{Park}},
  \bibinfo{journal}{Phys.Rev.Lett.} \textbf{\bibinfo{volume}{111}},
  \bibinfo{pages}{041801} (\bibinfo{year}{2013}), \eprint{1304.4936}.

\bibitem[{\citenamefont{Chen et~al.}(2014{\natexlab{a}})\citenamefont{Chen,
  Harnik, and Vega-Morales}}]{Chen:2014gka}
\bibinfo{author}{\bibfnamefont{Y.}~\bibnamefont{Chen}},
  \bibinfo{author}{\bibfnamefont{R.}~\bibnamefont{Harnik}}, \bibnamefont{and}
  \bibinfo{author}{\bibfnamefont{R.}~\bibnamefont{Vega-Morales}}
  (\bibinfo{year}{2014}{\natexlab{a}}), \eprint{1404.1336}.

\bibitem[{\citenamefont{Chen et~al.}(2014{\natexlab{b}})\citenamefont{Chen,
  Falkowski, Low, and Vega-Morales}}]{Chen:2014ona}
\bibinfo{author}{\bibfnamefont{Y.}~\bibnamefont{Chen}},
  \bibinfo{author}{\bibfnamefont{A.}~\bibnamefont{Falkowski}},
  \bibinfo{author}{\bibfnamefont{I.}~\bibnamefont{Low}}, \bibnamefont{and}
  \bibinfo{author}{\bibfnamefont{R.}~\bibnamefont{Vega-Morales}}
  (\bibinfo{year}{2014}{\natexlab{b}}), \eprint{1405.6723}.

\bibitem[{\citenamefont{Falkowski and Vega-Morales}(2014)}]{Falkowski:2014ffa}
\bibinfo{author}{\bibfnamefont{A.}~\bibnamefont{Falkowski}} \bibnamefont{and}
  \bibinfo{author}{\bibfnamefont{R.}~\bibnamefont{Vega-Morales}}
  (\bibinfo{year}{2014}), \eprint{1405.1095}.

\bibitem[{\citenamefont{Kanemura et~al.}(2014)\citenamefont{Kanemura, Kikuchi,
  Yagyu, and Yokoya}}]{Kanemura:2014goa}
\bibinfo{author}{\bibfnamefont{S.}~\bibnamefont{Kanemura}},
  \bibinfo{author}{\bibfnamefont{M.}~\bibnamefont{Kikuchi}},
  \bibinfo{author}{\bibfnamefont{K.}~\bibnamefont{Yagyu}}, \bibnamefont{and}
  \bibinfo{author}{\bibfnamefont{H.}~\bibnamefont{Yokoya}}
  (\bibinfo{year}{2014}), \eprint{1407.6547}.

\end{thebibliography}

\end{document}